\documentclass[a4paper, 12pt]{article}

\usepackage[english]{babel}
\usepackage[utf8x]{inputenc}
\usepackage[T1]{fontenc}

\usepackage[top=1.3cm, bottom=2.0cm, outer=2.5cm, inner=2.5cm, heightrounded,
marginparwidth=1.5cm, marginparsep=0.6cm, margin=2cm]{geometry}

\usepackage{graphicx} 
\usepackage[colorlinks=False]{hyperref} 
\usepackage{xcolor}
\usepackage{mathtools}
\usepackage{amsmath}  
\usepackage{amsfonts} %
\usepackage{amssymb}  %
\usepackage{authblk}
\usepackage{cite}

\makeatletter
\newcommand\appendix@section[1]{%
  \refstepcounter{section}%
  \orig@section*{Appendix \@Alph\c@section: #1}%
  \addcontentsline{toc}{section}{Appendix \@Alph\c@section: #1}%
}
\let\orig@section\section
\g@addto@macro\appendix{\let\section\appendix@section}
\makeatother

\newtheorem{theorem}{Theorem}
\usepackage{caption}
\usepackage{float}

\newcommand{\oL}{\overline{L}}
\newcommand{\oq}{\overline{q}}
\newcommand{\oh}{\overline{h}}

\newcommand{\mL}{\mathcal{L}}

\title{\bf High Energy Modular Bootstrap, \\ Global Symmetries and Defects}

\author[1]{Sridip Pal}
\author[2]{Zhengdi Sun}
\affil[1]{\it School of Natural Sciences, Institute for Advanced Study, Princeton, NJ 08540, U.S.A.}
\affil[2]{\it Department of Physics, 
University of California San Diego,
La Jolla, CA 92093, U.S.A.}

  \date{}
  
\begin{document}
\maketitle
\thispagestyle{empty}

\begin{abstract}
We derive Cardy-like formulas for the growth of operators in different sectors of unitary $2$ dimensional CFT in the presence of topological defect lines by putting an upper and lower bound on the number of states with scaling dimension in the interval $[\Delta-\delta,\Delta+\delta]$ for large $\Delta$ at fixed $\delta$. Consequently we prove that given any unitary modular invariant $2$D CFT symmetric under finite global symmetry $G$ (acting faithfully), all the irreducible representations of $G$ appear in the spectra of the untwisted sector; the growth of states is Cardy like and proportional to the ``square'' of the dimension of the irrep. In the Schwarzian limit, the result matches onto that of JT gravity with a bulk gauge theory. If the symmetry is non-anomalous, the result applies to any sector twisted by a group element. For $c>1$, the statements are true for Virasoro primaries. Furthermore, the results are applicable to large c CFTs. We also extend our results for the continuous $U(1)$ group. 
\end{abstract}

\clearpage

\tableofcontents
\pagenumbering{arabic} 
\clearpage

\section{Summary \& Discussion}

Symmetry plays a key role in studying Quantum Field theories (QFT). To study a QFT admitting a symmetry, we consider irreducible representations (irreps) of the group and declare that the quantum fields transform as irreps. A very natural and fundamental question is to ask whether there is any consistency condition telling us existence or absence of particular kind of irreps. These conditions can come about due to mathematical consistency and/or due to physical requirements like unitarity. Some of the famous examples in this genre are Coleman Mandula theorem \cite{Coleman:1967ad}, which roughly implies the impossibility of mixing space-time (Poincare) symmetry with internal symmetry unless one has supersymmetry; the unitarity bounds in $(3+1)$-D CFT by Mack \cite{Mack:1975je}, Weinberg-Witten theorem \cite{Weinberg:1980kq}, which shows that impossiblity of having massless particles with higher spin in a theory with Lorentz covariant energy momentum tensor/conserved current.

In this work, we consider unitary modular invariant $2$D conformal field theory. The consistency condition that we are going to leverage is modular transformation properties of Torus partition function with/without possible insertion of some operators.  A standard result along this line is the existence of infinite number of Virasoro primaries for $c>1$ CFTs \cite{cardy1986operator, Baur}. Recently, it has been established that every integer spin has to appear in the bosonic CFT \cite{Mukhametzhanov:2020swe} by projecting the grand canonical partition function of $2$D CFT onto a particular spin and studying the high temperature behavior of this fixed spin partition function. In a similar spirit, to study the different sectors of a $2$D CFT with global symmetries (more generically with insertion of topological defect lines), we project the canonical partition function (with/without the insertion of topological defect lines) onto relevant sectors and study the high temperature behavior to extract the growth of operators within each sector. Further use of modular crossing equations can be found in \cite{KM, Cardy:2017qhl, HKS, Das:2017cnv,Hikida:2018khg,Romero-Bermudez:2018dim,Brehm:2018ipf,Benjamin:2019stq,Alday:2019vdr,Collier:2019weq, Brehm:2019pcx, Baur, Ganguly:2019ksp, Pal:2019yhz, Pal:2019zzr, Mukhametzhanov:2020swe} and some aspects has been made symmetry sensitive in \cite{dattadaspal, Dyer:2017rul,Bae:2018qym, Lin:2019kpn}. 

One of the motivations for undertaking such investigation stems from a related question in holography. In the context of AdS-CFT, it is widely believed that all the irreps of internal gauge group appears in the gravity side a.k.a ``completeness hypothesis''; on the CFT side, the gauge symmetry becomes a global symmetry and hence it implies the existence of all the irreps of the global symmetry modulo some fine prints\cite{Polchinski:2003bq,Banks:2010zn,Harlow:2015lma,Harlow:2018jwu, Harlow:2018tng}. To understand it better, consider the case of $U(1)$. If we know that an operator with minimal charge exists, we can create black holes of arbitrary charge by collapsing such minimal charged objects in arbitrary number. On the CFT side, this amounts to taking OPE and generating operators of arbitrary charge. One of the main challenges is to show that the such minimal charged object exists, i.e. $U(1)$ acts faithfully. Here we will not be saying anything about faithfulness. Rather given the faithfulness condition on the CFT side, we will pose the following question of whether one can generate operators of arbitrary charge with arbitrarily high dimension in the way mentioned above. On the gravity side this amounts to having black holes with arbitrary charge. By the OPE argument, one can generate primaries of arbitrary charge and one needs to consider heavy descendants to answer positively to the above question. Hence, a more refined and nontrivial question is to ask whether we can say anything about heavy primaries with arbitrary charge and if possible, whether we can estimate the growth of each irreps. It turns out that in a $2$D CFT, one can investigate this leveraging the modular invariance.

 The recent study of partition function of $2$-D JT gravity \cite{Saad:2019lba} with bulk gauge field \cite{Kapec:2019ecr,Iliesiu:2019lfc} motivates us as well. It is well appreciated that the genus zero contribution to the partition function can be obtained by looking at dual quantum mechanical system, which is known to be the Schwarzian limit of a $2$-D CFT\cite{Mertens:2017mtv,Ghosh:2019rcj}. Now considering a bulk gauge field amounts to having a CFT with a global symmetry and then taking the Schwarzian limit. One curious feature present in the calculation of \cite{Kapec:2019ecr} is the square of dimension of irrep in the expression for density of states corresponding to the genus zero partition function of JT gravity with bulk gauge field. Here we take up a CFT calculation to precisely reproduce this curious factor.\\

Given a continuous global symmetry, we can turn on fugacity corresponding to the conserved current and consider the grand canonical partition function. This idea can be generalized to discrete symmetries by thinking of inserting topological defect lines (TDL) while doing the path integral over the relevant manifold to define the grand canonical partition function. In fact, one can allow non invertible TDLs (which does not correspond any global symmetry in conventional sense, nonetheless meaningful object, see section $1$ of \cite{Lin:2019hks}) and define grand canonical partition functions. In this work, the relevant manifold is square torus, i.e.\!~we consider $2$D CFT on a spatial circle of length $2\pi$, at inverse temperature $\beta$. If the topological defect line is inserted along the spatial circle, it is exactly the grand canonical partition function. If the topological line is inserted along the temporal circle, it creates a defect in the spatial manifold, thereby defines a ``defect'' Hilbert space of operators. The partition function constructed out of operators in the defect Hilbert space is related to the grand canonical partition function by a $S$ modular transformation. Roughly speaking, a $S$ modular transformation exchanges the spatial and temporal circle, thereby changes the role of TDLs. Given this set up, we ask following questions: 

\begin{itemize}
\item Can we estimate the growth of operators in the defect Hilbert space ? The spectrum of operators in the defect Hilbert space is not same as the original Hilbert space. On the other hand, one might think that introducing a defect only modifies the theory globally by modifying the boundary conditions of the field, thus one should not expect any change in asymptotic growth of operators compared to the original Hilbert space. We will confirm this intuition in part by doing a rigorous calculation in this work. Even though, the spectrum changes, the averaged behavior remains same (apart from a possible multiplicative factor, which we explain below in the results) even in the presence of defects. 

\item Given a $2$D CFT with a global symmetry (finite group), do all the irreps of the global symmetry  group appear in the spectra of local operators? The answer turns out to be yes.

\item If the symmetry group is non-anomalous, it is possible to group the operators appearing in the defect Hilbert space into irreps of the group and we ask whether all the irreps of the global symmetry  group appear in the defect spectra. Here also the answer turns out to be yes.
\end{itemize}

The basic strategy that we follow to answer these questions is to consider a partition function of the the sector of the CFT which we want to study and then to look at its high temperature behavior. The relevant sector specific partition function can be obtained by using appropriate projection operators onto the partition function in appropriate channel. The precise way of doing this is explained in details in the paper.  Below we summarize our results and discuss the implications. 

\paragraph{Results:}
\begin{enumerate}
\item We consider a CFT on a torus with the topological defect line (TDL) being inserted along the temporal direction. We estimate the growth of operators in the defect Hilbert space $\mathcal{H}_{\mathcal{L}}$ as $\Delta\to\infty$ :
\begin{equation}
\text{growth of operators in}\ \mathcal{H}_{\mathcal{L}}\simeq  N_0\rho_0(\Delta)
\end{equation}
where 
\begin{equation}\label{def:rho0}
\rho_0(\Delta)=\left(\frac{c}{48\Delta^3}\right)^{\frac{1}{4}}\exp\left[2\pi\sqrt{\frac{c\Delta}{3}}\right]\,.
\end{equation}
A rigorous statement is made in theorem~[\ref{th:defectcount}] and in the  eq.~\eqref{eq:defectcount}. The TDL can either correspond to a Global symmetry or correspond to a non-invertible defect such as duality defect. Here $N_0$ is ``quantum dimension'', obtained from the action of TDL on the $\Delta=0$ state. For TDL corresponding to global symmetry the vacuum remains invariant and $N_0=1$, for TDL corresponding to duality defects, $N_0$ may not be $1$. 

\item We consider a CFT with a finite global symmetry group (acting faithfully). We find that every irreducible representation has to appear in the spectrum of operators in the untwisted sector and they have a Cardy like growth as $\Delta\to\infty$. In particular, we have \begin{equation}
\begin{aligned}
&\text{growth of occurence of particular irrep}\  \alpha \simeq d_{\alpha}|G|^{-1}\rho_0(\Delta)\,,\\
&\text{growth of states in an irrep}\ \alpha \simeq d_{\alpha}^2|G|^{-1}\rho_0(\Delta)\,.
\end{aligned}
\end{equation} 
where $\rho_0(\Delta)$ is defined in eq.~\eqref{def:rho0}. Here $|G|$ is the order of the group and $d_{\alpha}$ is the dimension of the representation of irrep $\alpha$. We remark that if we sum over all the irreps, we get back the usual Cardy like growth for all the operators, i.e.\!~$\rho_0(\Delta)$ . A more rigourous statement is made in theorem~[\ref{th:irrepcount}] and in eq.~\eqref{eq:irrepcount}. If the symmetry is non-anomalous, the result is true for any particular twisted sector. The rigorous statement can be found in theorem~[\ref{th:irrepcounttw}]. To illustrate, in the example of $\mathbb{Z}_2$, $\alpha$ can be even or odd, $d_{\alpha}=1$ and $|G|=2$.

$\bigstar$ A unified version of the above two results is presented in theorem~[\ref{th:vvm}] and in eq.~\eqref{eq:vvm}.

$\bigstar$ {\it Schwarzian sector-JT gravity}: $2$-D CFT is known to have a schwarzian sector \cite{Ghosh:2019rcj}, which is relevant for the study of JT gravity. The partition function corresponding to the disk topology \cite{Saad:2019lba} corresponds to the identity character in some particular limit, as explained in \cite{Mertens:2017mtv,Ghosh:2019rcj}. Having a global symmetry on the CFT side induces a bulk gauge field on the gravity side. In the set up \cite{Kapec:2019ecr}, the bulk gauge theory is taken to be topological BF theory.  The corresponding partition function has been calculated in \cite{Kapec:2019ecr,Iliesiu:2019lfc},  the density of states has been shown to have a $d_{\alpha}^2|G|^{-1}$ factor multiplied with the seed gravity answer without the gauge field. Our result precisely reproduces this factor, since we can readily take the Schwarzian limit of our answer following \cite{Mertens:2017mtv,Ghosh:2019rcj}. 

$\bigstar$ The factor $d^2_{\alpha}$ as opposed to $d_{\alpha}$ might be surprising because we expect the extra $d_{\alpha}$-fold degeneracy due to the symmetry. Intuitively, this comes about due to smearing\footnote{We thank Raghu Mahajan for discussion along this line.}. The growth formula is valid only after smearing over an order one window, and it turns out that the order one window has $d_{\alpha}$ number of $\alpha$ irreps. This might hint at emergence of some approximate symmetry. This is exactly similar to the scenario in \cite{Kapec:2019ecr} where they speculate about emergence of extra approximate symmetry. We discuss this after \eqref{pointer} as one of the remarks and we explicitly look at 3-state Potts model ($c=4/5$)  to back up our claim.

$\bigstar$ The rigorous bounds in theorem~[\ref{th:defectcount}] and theorem~[\ref{th:irrepcount}] have order one error. Without any further input, that's the best order of error that one can achieve. To optimize over the order one error, we need to use Selberg-Beurling extremizers as elucidated in \cite{Mukhametzhanov:2020swe} .

\item All of the above estimates can be made in the limit $c\to\infty$ and $\Delta=c\left(\frac{1}{12}+\epsilon\right)$ for $\epsilon>\frac{1}{12}$ following \cite{HKS,Baur}. We use $\Delta-\frac{c}{12}$ instead of $\Delta$ everywhere in the above formulas.
\item All of the above estimates can be made sensitive to Virasoro primaries for $c>1$ following \cite{Baur}. Instead of $\rho_0(\Delta)$ we will have 
$$\rho_{0}^{\text{Vir}}(\Delta)=\left(\frac{c-1}{3}\Delta\right)^{\frac{1}{4}}\exp\left[2\pi\sqrt{\frac{(c-1)\Delta}{3}}\right]\,.$$
\item We find the analogous result for continuous group $U(1)$ (acting faithfully). Under a technical assumption, we show that every charged state has to appear in the spectrum and they do have a Cardy like growth at large $\Delta$ given by $\sqrt{\frac{c}{3k}}\frac{1}{\Delta}\exp\left[2\pi\sqrt{\frac{c\Delta}{3}}\right] $. The rigorous statement can be found in eq.~\eqref{eq:u1count}. Again one can generalize this to Virasoro primaries for $c>1$ in one hand and on the other hand to the large central charge regime.
\end{enumerate}

$\bullet$ {\bf Application and future avenues:}\\

\begin{figure}[H]
\centering
\includegraphics[scale = 0.25]{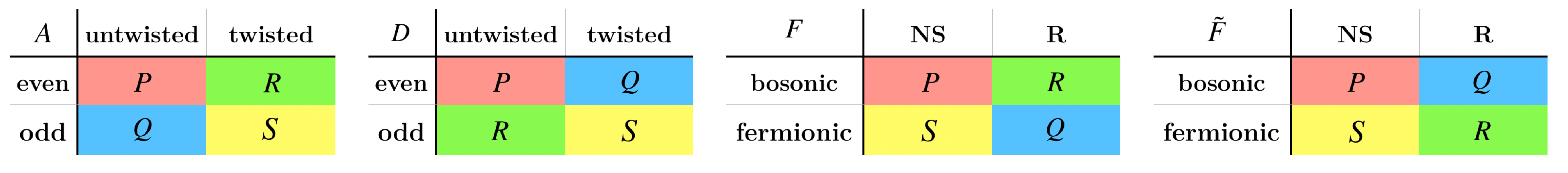}
\caption{The theory $A$ and $D$ are related by orbifolding by $\mathbb{Z}_2$. The theory $A$ and $F$ are related by Bosonization-Fermionization and so are $D$ and $\tilde{F}$.}
\label{table}
\end{figure}

We have already mentioned one application of our result upon taking the Schwarzian limit and making contact with the results of JT gravity with a bulk gauge field. Here we list out few more applications. For example, we can consider the following table~\ref{table}. A similar one appears in \cite{Hsieh:2020uwb}. Same is explored in the context of $\mathbb{Z}_2$ symmetry of Monster CFT in \cite{Lin:2019hks}. We consider a theory $A$ with a non-anomalous $\mathbb{Z}_2$. The untwisted sectors can be divided into two pieces: even and odd, named as $P$ and $Q$.  This is obtained when the TDL corresponding to the $\mathbb{Z}_2$ symmetry is extended along the spatial direction. The twisted sector is obtained by keeping the TDL along the time direction, thereby creating a defect. Since, $\mathbb{Z}_2$ is non-anomalous, one can have even and odd states in the twisted sector as well, we call them $R$ and $S$ respectively. Gauging this $\mathbb{Z}_2$ symmetry lands us onto the theory D.  Both the theory $A$ and $D$ can be fermionized to theory $F$ and $\tilde{F}$. The effect of this amounts to permuting and relabelling the different sectors $P,Q,R,S$. Using our result, we can estimate the growth of operators for each of the sector $P,Q,R,S$. All of them have a Cardy like growth given by $\frac{1}{2}\rho_0(\Delta)$ (corresponding to $d_{\alpha}=1$ and $|G|=2$) for large $\Delta$.

One can think of further applications of these ideas generalizing the results appearing in \cite{KM,Das:2017cnv,Hikida:2018khg,Romero-Bermudez:2018dim,Brehm:2018ipf}. Moreover, one can also make all of the above results spin-sensitive following \cite{Mukhametzhanov:2020swe}. It would also be interesting to explore other aspects of modular bootstrap for example bounding the dimension of lowest nontrivial Virasoro primary, constructing the extremal functionals \cite{Hellerman:2009bu,Collier:2016cls,Cho:2017fzo,Bae:2017kcl,Mazac:2018mdx,Mazac:2018ycv,Afkhami-Jeddi:2019zci,Hartman:2019pcd,Ashrafi:2019ebi} in presence of TDLs.\\

As a technique, we generalize the application of Tauberian formalism in context of CFT beyond $S$ modular invariant partition functions. In particular, the method can be applied to vector valued modular functions as elucidated in \S\ \ref{sec:vvm}. One immediate application would be generalizing the results of \cite{Das:2017cnv} for $LL'H$-squared for two different operators using the Tauberian technique. Note that the positivity is guaranteed in one of the channels while in other channel, it is not there. This scenario is reminiscent of the partition function of the defect Hilbert space, where positivity is guaranteed but in the $S$ transformed channel, positivity is not guaranteed.\\

$\bullet$ {\bf Organization}\\

The paper is organized in following manner. The \S\ \ref{review} reviews the idea of TDLs as generalization of global symmetry. A nice and brief exposition can also be found in the introduction of \cite{Lin:2019hks}. In \S\ \ref{sec:defect}, we study the defect Hilbert space. In \S\ \ref{sec:irrep} we study the growth of operators within an irrep. The \S\ \ref{subsec:warm} expounds on a simple example of $\mathbb{Z}_2$ symmetry, which we generalize and make rigorous in \S\ \ref{subsec:general}. The similar question relevant to $U(1)$ symmetry is analyzed in \S\ \ref{subsec:u1}. The \S\ \ref{sec:vvm} encapsulates the gist of applying the Tauberian technique to the vector valued modular functions. In \S\ \ref{sec:verify}, we provide some numerics on known models to cross-check our results. In \S\ \ref{sec:spin}, we review the derivation of spin selection rule for anomalous global symmetry.

\section{Lightning review of Topological defect line}\label{review}

Given any continuous global symmetry, one can define Noether's current $j_{\mu}$ and the charge $Q$ is given by $Q=\int \text{d}^{d-1}x\ j^{0}$, an integral of $j_{\mu}$ over a codimension one surface, here the surface is given by $x^{0}=\text{constant}$. In general, one can define an operator, supported on any codim-$1$ surface $\Sigma$ and given by $\exp(\imath \theta \int_\Sigma \star j)$.  The statement that the charge conservation, $\text{d}\!\star\!j = 0$ boils down to the statement that the operator is invariant under continuous small deformation $\Sigma$. We also note that here the charge $Q$ is a scalar, we name it $0$ form symmetry. Now instead of codimension $1$ surface, one can in general consider topological surface operator of codim-$(q+1)$ and define $q$-form global symmetries \cite{Gaiotto:2014kfa}.  For a $0$ form symmetry, when the surface $\Sigma$ is chosen to be the full spatial slice, this operator is exactly the symmetry operator acting on the Hilbert space; while if one of the direction of $\Sigma$ is the time direction, then this operator creates a codim-$1$ defect in the space--any local operator undergoes a twisting when crossing the defect. For this reason, topological surface operators are sometimes called the topological defects. 

In 2-dim, ordinary $0$-form symmetries correspond to topological defects lines (TDL). A natural question to ask if whether converse is true. The answer is generically no for the following reason. The fusion of the TDLs associated with global symmetries must respect the group multiplication. Therefore, for any TDL corresponding to an group, there must exist an inverse TDL; in fact, the inverse line can be obtained by simply reversing the orientation of the line. However, there do exist the so-called non-invertible line operators which don't have an inverse, (e.g. the duality line $N$ in the Ising CFT or Monster CFT \cite{Lin:2019kpn,Lin:2019hks}). 

As in the general dimension space-time, we can place the TDL $\mathcal{L}$ along the time direction on $\mathbb{R}_t\times S_1$, which amounts to imposing the twisted boundary condition on $S_1$. The resulting Hilbert space is called the defect Hilbert space $\mathcal{H}_{\mathcal{L}}$ whose states can be labelled by the usual weights $(h,\oh)$. This is possible because the energy momentum tensor commutes with TDL. Via state-operator correspondence, a state in $\mathcal{H}$ corresponds to an operator, sitting at the end of the $\mathcal{L}$. A particular important question for our analysis is whether there's a state with conformal weight $(0,0)$ in the defect Hilbert space. As in \cite{Lin:2019kpn}, if we require that the global symmetry acts faithfully on the Hilbert space of local operators, that is, the only line operator that commutes with every local operators is the identity line, then the defect Hilbert space $\mathcal{H}_{\mathcal{L}}$ contains no weight-$(0,0)$ state. Otherwise, the existence of such state would allow line $\mathcal{L}$ to connect to the identity line via the corresponding operator, thus it would commute with every local operator, violating our requirement (see fig.~\ref{No-weight-0-state}). As we will see, this makes sure the leading result in our analysis is universal in the sense that it only depends on the central charge $c$ and the symmetry group $G$.  We also remark here if the symmetry is anomalous (if one can not define action of the symmetry in the defect Hilbert space consistently), then the ground state in the defect Hilbert space has $\Delta>0$. This follows from the spin selection rule \cite{Lin:2019kpn}. We review it in the appendix~\S\ref{sec:spin}.

\begin{figure}[H]\label{No-weight-0-state}
\centering
\includegraphics[scale = 0.3]{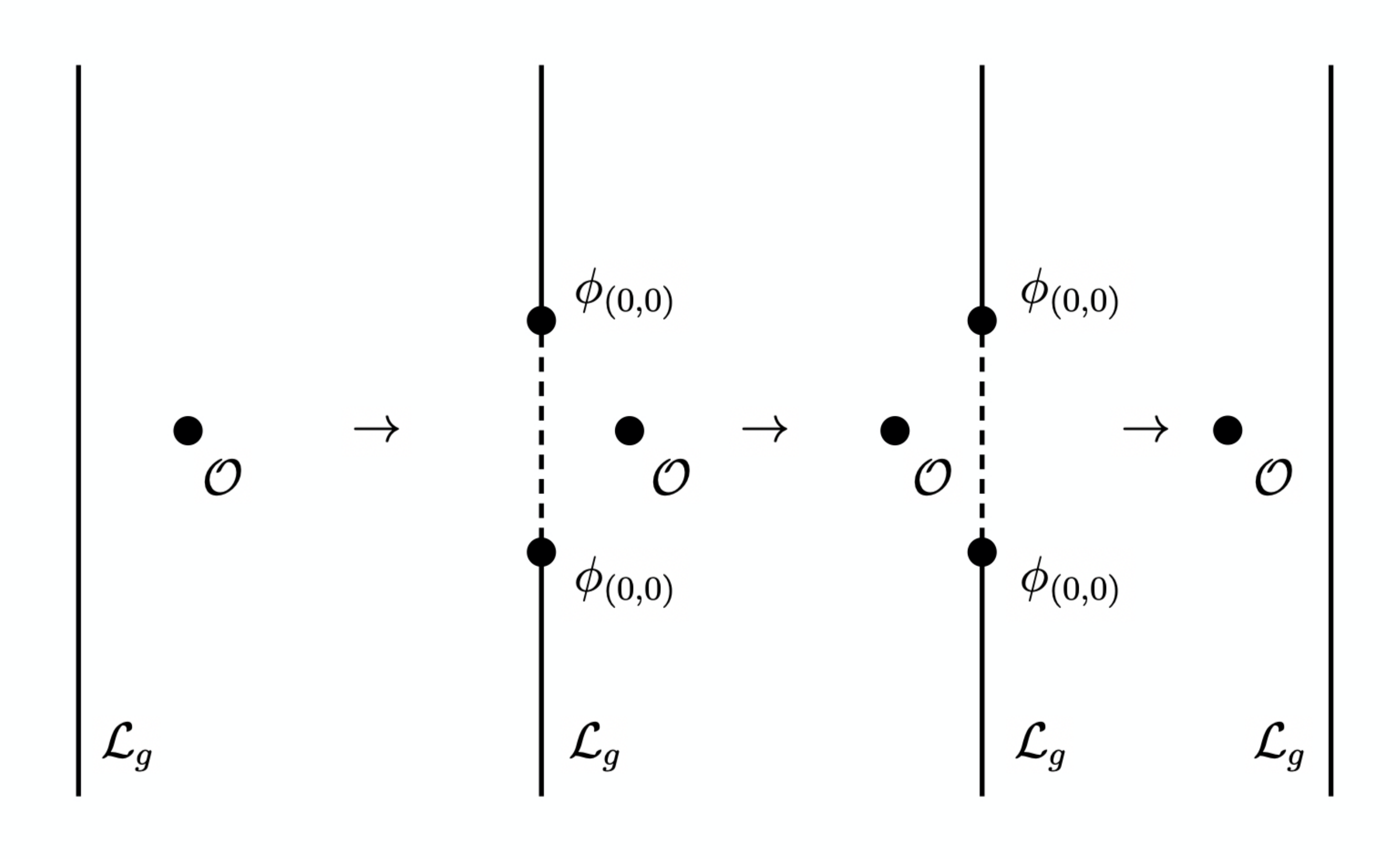}
\caption{$\mathcal{L}_g$ denotes the $g$ symmetry TDL; dashed line denotes the trivial line; $\mathcal{O}$ is arbitrary local operator; and $\phi_{(0,0)}$ is the operator correspond to the weight $(0,0)$ state in $\mathcal{H}_{\mathcal{L}_g}$. The existence of $\phi_{(0,0)}$ allows us to open the TDL to show that the $\mathcal{L}_g$ commutes with every local operator $\mathcal{O}$. In the main text, we drop the $g$ subscript where it is redundant.}
\label{No-weight-0-state}
\end{figure}

\begin{figure}[H]
\centering
\includegraphics[scale = 0.3]{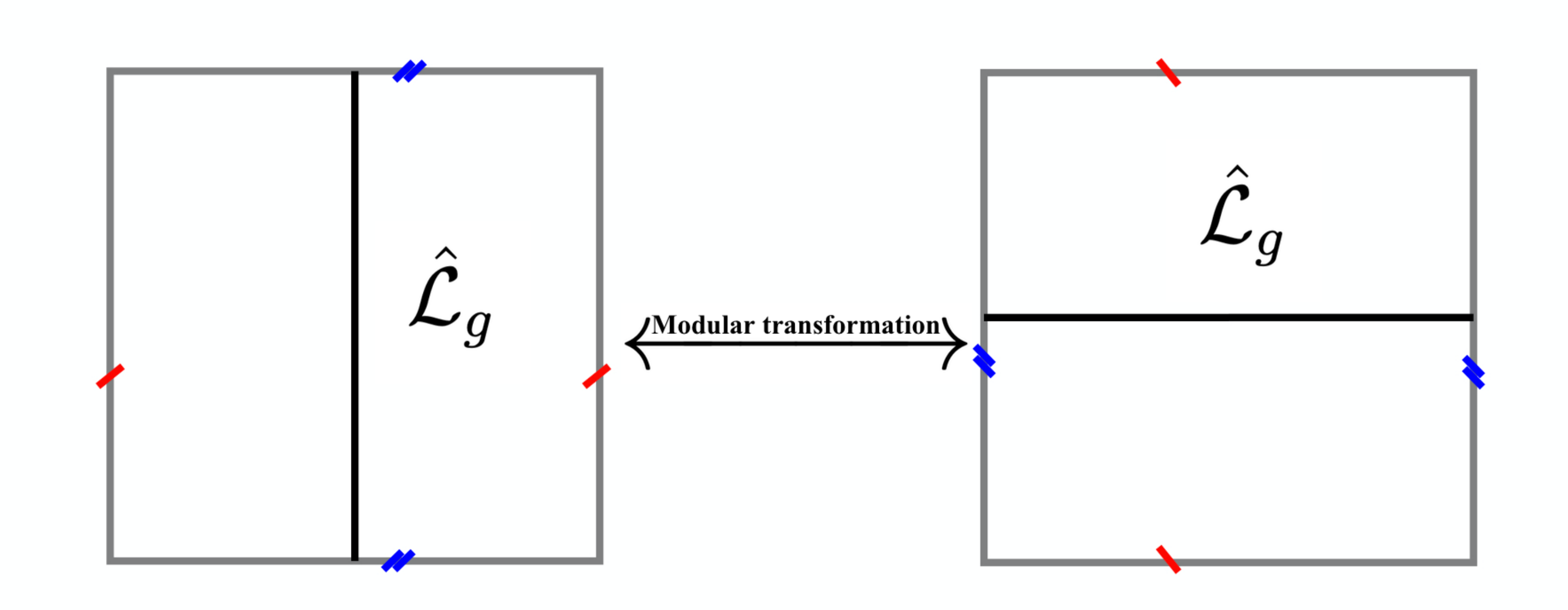}
\caption{The partition function of the defect Hilbert space (the left figure, which we will denote $Z_{\mathcal{L}}(\beta,g)$) is related to the partition function with the insertion of the corresponding charge operator (the right figure, which we will denote $Z^{\mathcal{L}}(\beta,g)$).}
\label{modular-transformation}
\end{figure}

On the other hand, if we place the TDL along the spatial direction, then it acts as a group element on the Hilbert space of local operators. Instead of $\mathbb{R}_t\times S_1$, one can consider $S_1\times S_1$ and generalize the above story. Since the modular transformation exchanges two cycles of $S_1\times S_1$, the configuration of TDL along the spatial circle must be related to the configuration of TDL along the temporal circle. This brings us to the key property of the partition function of defect Hilbert space, that is, it is related to the partition function with the insertion of the corresponding charge operator (see fig.~\ref{modular-transformation}) along the spatial cycle. To be concrete, we define
\begin{equation}
\begin{aligned}
	Z_{\mathcal{L}}(\beta,g) &:= \text{Tr}_{\mathcal{H}_{\mathcal{L}_g}}(q^{L_0-c/24}\oq^{\oL-c/24}), \\
	Z^{\mathcal{L}}(\beta,g) &:= \text{Tr}_{\mathcal{H}}(\hat{g}\ q^{L_0-c/24} \oq^{\oL-c/24}), \\
\end{aligned} 
\end{equation}
and modular transformation tells us that
\begin{equation}
	Z_{\mathcal{L}}(\beta,g) = Z^{\mathcal{L}}\bigg(\frac{4\pi^2}{\beta},g\bigg).	
\end{equation}

We end this section by making a crucial remark that the low temperature expansion coefficient of $Z_{\mathcal{L}}(\beta,g)$ is positive, hence falls under the purview of Tauberian formalism whereas in the dual channel, positivity is not guaranteed. One needs to keep this in mind while expecting whether a Cardy like statement is true or not. For example, whereas we can hope to prove the asymptotic growth of low temperature expansion coefficient of $Z_{\mathcal{L}}(\beta,g)$, the same is not true for $Z^{\mathcal{L}}(\beta,g)$ without any further assumption because the positivity is not guaranteed in this channel.

\section{Charting Defect Hilbert Space $\mathcal{H}_{\mL}$ associated with TDL $\mathcal{L}$}\label{sec:defect}

In the usual Cardy formula, the asymptotic growth of operators is controlled by the low temperature limit of the partition function in the dual ($S$ transformed) channel. As explained in the previous section \S\ \ref{review}, the dual channel corresponding to the partition function of a defect Hilbert space ($Z_{\mathcal{L}}(\beta,g)$) is the partition function evaluated on the original Hilbert space with an insertion of group element $g$, which we call $Z^{\mathcal{L}}\left(\frac{4\pi^2}{\beta} ,g\right) $. The leading behavior (low temperature) in the later channel is controlled by the vacuum operator. Thus one can expect a Cardy like growth for operators in the original channel i.e in the defect Hilbert space. 
\begin{equation}
Z_{\mathcal{L}}(\beta \to 0,g) = Z^{\mathcal{L}}\left(\frac{4\pi^2}{\beta} \to 0,g\right) \simeq e^{\frac{\pi^2 c}{3\beta}}
\end{equation}
and hence we expect the growth of the operators in the defect Hilbert space is given by inverse Laplace of $e^{\frac{\pi^2 c}{3\beta}}$, which is $\rho_0(\Delta)$. In what follows, we will be making this idea rigorous using Tauberian techniques.

\subsection{Cardy Formula for Defect Hilbert Space}

\begin{theorem}\label{th:defectcount}
Given a TDL $\mathcal{L}$, the asymptotic behavior ($\Delta\to \infty$) of the growth of states in an order one window of width $2\delta$, centered at $\Delta$ in the defect Hilbert space $\mathcal{H}_{\mathcal{L}}$ is given by
\begin{equation}\label{eq:defectcount}
\begin{aligned}
&c_-N_0\rho_0(\Delta)\leq \frac{1}{2\delta}\int_{\Delta-\delta}^{\Delta+\delta}\text{d}\Delta'\ \rho_{\mathcal{H}_{\mathcal{L}}}(\Delta')\leq c_+N_0\rho_0(\Delta)
\end{aligned}
\end{equation}
where $N_0=1$ if the TDL is associated with a global symmetry, i.e.\!~invertible one, otherwise it is taken to be a positive number as defined below and $\rho_0(\Delta)$ is defined in eq.~\eqref{def:rho0}. Here $c_{\pm}$ order one positive numbers. These numbers can be determined using the extremal functionals appearing in \cite{Mukhametzhanov:2020swe}. In particular, we have $c_{\pm}=1\pm 1/2\delta$. The above statement is true under the following technical assumptions:
\begin{itemize}
	\item The action of $\mathcal{L}$ on the states are uniformly bounded, i.e.\!~$|\langle\Delta|\mathcal{L}|\Delta\rangle| \leq N$ for all $\Delta$ in the physical spectra. For example, if we consider $\mathbb{Z}_{n}$ then, $|\langle\Delta|\mathcal{L}|\Delta\rangle| \leq 1$, since the matrix element is always a phase. In fact, this is true for any TDL associated with a finite group. For non invertible TDLs, i.e.\!~the ones which are not associated with global symmetry, we take this as an assumption, which is true for a wide class of non invertible TDLs.
	\item The vacuum is invariant under any topological defect line associated with global symmetry. Thus we have
	\begin{equation}
		\mathcal{L}|0\rangle = |0\rangle,
	\end{equation}
	\item The action of a non-invertible topological defect line $\mathcal{L}$ (such as duality defects, not associated with any global symmetry) on the vacuum state is given by:
	\begin{equation}
		\mathcal{L}|0\rangle = N_0 |0\rangle, \,\,\, \, N_0 > 0.	
	\end{equation}
	For example, in Ising model, we have duality defect line $\widehat{N}$ and $\widehat{N}|0\rangle= \sqrt{2}|0\rangle$.
\end{itemize}
\end{theorem}

The basic structure of the proof is similar to the one appeared in \cite{Baur, Pal:2019zzr, Mukhametzhanov:2020swe}, though the deatils are different as we will see. This comment applies to theorems proven in subsequent sections as well. We start by considering two functions $\phi_{\pm}(\Delta)$ whose Fourier transformation has finite support $[-\Lambda,\Lambda]$ and they majorise and minorise the characteristic function for the interval $[\Delta-\delta,\Delta+\delta]$:
\begin{equation}
	\phi_- (\Delta') \leq \theta_{[\Delta - \delta,\Delta + \delta]}(\Delta') \leq \phi_+(\Delta')\,.	
\end{equation}
From the above it follows that
\begin{equation}
	e^{\beta(\Delta-\delta)}e^{-\beta \Delta'}\phi_-(\Delta') \leq \theta_{[\Delta-\delta,\Delta + \delta]}(\Delta') \leq e^{\beta(\Delta+\delta)}e^{-\beta\Delta'}\phi_+(\Delta')\,.
\end{equation}
Multiplying both sides by the density of states of the twisted Hilbert space $\rho_{\mathcal{H}_\mathcal{L}}$ and integrating from $0$ to $\infty$, we find
\begin{equation}
	e^{\beta(\Delta-\delta)}\int_0^\infty dF(\Delta') e^{-\beta\Delta'}\phi_-(\Delta') \leq \int_{\Delta - \delta}^{\Delta + \delta} dF(\Delta') \leq e^{\beta(\Delta+\delta)}\int_0^\infty dF(\Delta') e^{-\beta\Delta'}\phi_+(\Delta')\,,	
\end{equation}
where $dF(\Delta') = \rho_{\mathcal{H}_{\mathcal{L}}}(\Delta') d\Delta'$. We emphasize $\beta,\delta$ are free parameters.  We consider the Fourier transformation of $\phi_\pm(\Delta) = \int_{-\infty}^{\infty} \text{d}t\ \hat{\phi}_{\pm}(t)^{-\imath\Delta t}$, such that in Fourier domain the above inequality becomes
\begin{equation}
\begin{aligned}
	&e^{\beta(\Delta-\delta)} \int_{-\infty}^{\infty}dt \, \hat{\phi}_-(t)Z_{\mathcal{L}}(\beta + \imath t) e^{-(\beta + \imath t) c/12}\\
	& \leq \int_{\Delta - \delta}^{\Delta+\delta} dF(\Delta')\leq \\
	& e^{\beta(\Delta + \delta)}\int_{-\infty}^{\infty} dt \, \hat{\phi}_+(t)Z_{\mathcal{L}}(\beta+\imath t) e^{- (\beta + \imath t) c/12}.	
	\end{aligned}
\end{equation}
The modular property implies
\begin{equation}
	Z_{\mathcal{L}}(\beta + \imath t) = Z^{\mathcal{L}}\bigg(\frac{4\pi^2}{\beta + \imath t}\bigg).
\end{equation}
Thus in the dual channel we have an expression in terms of the original Hilbert space. We split this original Hilbert space $\mathcal{H}$ into light part and heavy part: 
\begin{equation}
	Z^{\mathcal{L}}\bigg(\frac{4\pi^2}{\beta + \imath t}\bigg) = Z^{\mathcal{L}}_L\bigg(\frac{4\pi^2}{\beta + \imath t}\bigg) + Z^{\mathcal{L}}_H\bigg(\frac{4\pi^2}{\beta + \imath t}\bigg).
\end{equation}

Notice that the contribution from the light sector $Z_L^{\mathcal{L}}$ is not necessary real if it contains operators arbitrarily charged under global symmetry group $G$. For example, if we consider the $\mathbb{Z}_3$ symmetry, then the TDL $\mathcal{L}$ can act on a state such the state picks up a phase of $e^{2\pi \imath/3}$. One can circumnavigate this by assuming charge conjugation invariance. 
\begin{equation}
	\left(\int_{-\infty}^{\infty} dt \, \hat{\phi}_{\pm}(t) Z_{\mathcal{L}}(\beta + \imath t) e^{-(\beta + \imath t)c/12}\right)=\left(\int_{-\infty}^{\infty} dt \, \hat{\phi}_{\pm}(t) Z^{\mathcal{L}}\bigg(\frac{4\pi^2}{\beta + \imath t}\bigg)  e^{-(\beta + \imath t)c/12}\right) \in\mathbb{R} 
\end{equation}
Then we can split it as
\begin{equation}
\begin{aligned}
	&  \, \int_{-\infty}^{\infty} \text{d}t \, \hat{\phi}_{\pm}(t) Z^{\mathcal{L}}\bigg(\frac{4\pi^2}{\beta + \imath t}\bigg) e^{-(\beta + \imath t)c/12} \\
	= & \int_{-\infty}^{\infty} \text{d}t \, \hat{\phi}_{\pm}(t) Z^{\mathcal{L}}_L\bigg(\frac{4\pi^2}{\beta + \imath t}\bigg) e^{-(\beta + \imath t)c/12}+\int_{-\infty}^{\infty} dt \, \hat{\phi}_{\pm}(t) Z^{\mathcal{L}}_H\bigg(\frac{4\pi^2}{\beta + \imath t}\bigg) e^{-(\beta + \imath t)c/12}.
\end{aligned}		 
\end{equation}

At first, we consider the light sector where $\Delta \leq c/12$ choose a $\rho_0^{\mathcal{L}}(\Delta)$ such that 
\begin{equation}
	\int_0^\infty \text{d}\Delta\ \rho_0^{\mathcal{L}}(\Delta) e^{-\beta(\Delta-c/12)} = Z_L^{\mathcal{L}}\bigg(\frac{4\pi^2}{\beta}\bigg)\,.
\end{equation}
As a result, the contribution from the light sector can be written as 
\begin{equation}
\begin{aligned}
	& e^{\beta(\Delta \pm \delta)} \, \int_{-\infty}^{\infty} \text{d}t\ \hat{\phi}_{\pm}(t) Z_L^{\mathcal{L}}\bigg(\frac{4\pi^2}{\beta + \imath t}\bigg) e^{-(\beta+\imath t)c/12} = e^{\beta(\Delta \pm \delta)}  \, \int_0^\infty \text{d}\Delta'\ \rho_0^{\mathcal{L}}(\Delta') \phi_{\pm}(\Delta') e^{-\beta \Delta'}.
\end{aligned}
\end{equation}

Notice that, in general, the light sector contains all the states with $\Delta \leq c/12$. $\rho_0^{\mathcal{L}}(\Delta)$ contains more than just contribution from the vacuum state $N_0 \rho_0(\Delta)$ where $\rho_0(\Delta)$ is the crossing kernel of the vacuum state. The extra light operators would give exponentially suppressed corrections and are not universal (model dependent). Since there are finite number of operators below $c/12$, so that sum of the contribution coming from each of the extra light operators is still suppressed. In the following, we shall only consider the vacuum contribution.\\

Next, we treat contribution from the heavy sector and show they are suppressed in magnitude, hence can be dropped from both the lower and the upper bound.
\begin{equation}
\begin{aligned}\label{suppress}
 & e^{\beta(\Delta\pm\delta)} \bigg|\int_{-\infty}^{\infty} dt \, \hat{\phi}_{\pm}(t) Z^{\mathcal{L}}_H\bigg(\frac{4\pi^2}{\beta + \imath t}\bigg) e^{-(\beta + \imath t)c/12} \bigg| 
	\leq e^{\beta(\Delta - c/12 \pm \delta)} \int_{-\infty}^{\infty}dt \, \bigg|Z_{H}^{\mathcal{L}}\bigg(\frac{4\pi^2}{\beta + \imath t}\bigg)\bigg| |\hat{\phi}_{\pm}(t)| 
\end{aligned}
\end{equation}
Now we do the following estimation
\begin{equation}
\begin{aligned}
	\bigg|Z_{H}^{\mathcal{L}}\bigg(\frac{4\pi^2}{\beta+\imath t}\bigg)\bigg| &= \bigg|\sum_{\Delta > c/12} N_\Delta \exp\bigg[-\frac{4\pi^2}{\beta+\imath t}\bigg(\Delta-\frac{c}{12}\bigg)\bigg]\bigg| \\
	&\leq N \sum_{\Delta > c/12}\exp\bigg[-\frac{4\pi^2\beta}{\beta^2+t^2}\bigg(\Delta-\frac{c}{12}\bigg)\bigg] \\
	&= N Z_{H}\bigg[\frac{4\pi^2\beta}{\beta^2+t^2}\bigg] \leq N Z_{H}\bigg[\frac{4\pi^2\beta}{\beta^2+\Lambda^2_\pm}\bigg] \,\,\, \text{for} \,\,\, t^2 \leq \Lambda_{\pm}^2
\end{aligned}
\end{equation}
where $N_\Delta$ denote the action of $\mathcal{L}$ on a state with conformal dimension $\Delta$ and $N$ denote the upper bound of all $N_\Delta$'s. We use this bound in \eqref{suppress} and the fact that $\hat{\phi}_\pm(t)$ has finite support $[-\Lambda_\pm,\Lambda_\pm]$ to have the following inequality
\begin{equation}
\begin{aligned}
	& e^{\beta(\Delta-\delta)}\bigg[\,\int_0^\infty d\Delta' \, \rho_0^{\mathcal{L}}(\Delta') e^{-\beta \Delta'}\phi_-(\Delta') - N e^{-\beta c/12}Z_H\bigg(\frac{4\pi^2\beta}{\beta^2 + \Lambda_-^2}\bigg) \int_{-\Lambda_-}^{\Lambda_-} dt \, |\hat{\phi}_-(t)| \bigg] \\
	& \leq \int_{\Delta - \delta}^{\Delta + \delta} dF(\Delta') \leq \\
	& e^{\beta(\Delta+\delta)}\bigg[\,\int_0^\infty d\Delta' \, \rho_0^{\mathcal{L}}(\Delta') e^{-\beta \Delta'}\phi_+(\Delta') + N e^{-\beta c/12}Z_H\bigg(\frac{4\pi^2\beta}{\beta^2 + \Lambda_+^2}\bigg) \int_{-\Lambda_+}^{\Lambda_+} dt \, |\hat{\phi}_+(t)| \bigg].
\end{aligned}	
\end{equation}

The bounds get greatly simplified once we consider the large $\Delta$ region. Indeed, as in \cite{Baur} using HKS bound, one can show 
\begin{equation}
	e^{\beta \Delta}Z_{H}\bigg(\frac{4\pi^2\beta}{\beta^2 + \Lambda_{\pm}^2}\bigg) \sim e^{\beta \Delta} e^{\frac{\pi^2 c}{3\beta} \big(\frac{\Lambda_{\pm}}{2\pi}\big)^2} \sim \rho^{\mathcal{L}}_{0}(\Delta)^{1 + \frac{1}{2}\big(\big(\frac{\Lambda_{\pm}^2}{2\pi}\big) - 1\big)},
\end{equation}
where we choose $\beta = \pi\sqrt{\frac{c}{3\Delta}} <<1$. Therefore the contribution from $Z_H$ is sub-leading once we choose $\Lambda_{\pm} < 2\pi$. Then the upper bound at large $\Delta$ (the lower bound is similar $\phi_{+}\to\phi_{-}$) simplifies to 
\begin{equation}
\begin{aligned}
\int_{\Delta-\delta}^{\Delta+\delta} \text{d}\Delta' \ \rho_{\mathcal{H}_{\mathcal{L}}}(\Delta')  \leq e^{\beta\Delta} \int_0^\infty d\Delta' \, N_0 \rho_0(\Delta') \phi_+(\Delta') e^{-\beta\Delta'}\,.
\end{aligned}	
\end{equation}
Upon doing integrals by the saddle point approximation, we have in the $\Delta\to\infty$ limit
\begin{equation}
	N_0 c_- \rho_0(\Delta)\leq \frac{1}{2\delta} \int_{\Delta-\delta}^{\Delta + \delta} dF(\Delta') \leq N_0 c_+ \rho_0 (\Delta)\,, \quad \text{where}\ c_{\pm}=\frac{\pi}{\delta}\hat{\phi}_{\pm}(0)\,.
\end{equation}

This concludes the proof of the theorem. For $c>1$ CFTs, the analysis can be made sensitive to primaries only. We end this subsection with two remarks.

\begin{itemize}
\item As in \cite{Ganguly:2019ksp,Mukhametzhanov:2020swe}, we can derive a spectral gap for the defect Hilbert $\mathcal{H}_{\mathcal{L}}$. The upper bound on the gap is found to be $1$. This is the optimal gap as one can consider the Monster CFT with insertion of Identity line; now the defect Hilbert space is same as the original Hilbert space, as a result the gap is exactly $1$. For further discussion related to optimality, we refer the readers to \cite{Mukhametzhanov:2020swe}.

\item Following \cite{Baur}, we can also derive a global approximation of the number of states $F_{\mathcal{L}}(\Delta)$ in the defect Hilbert space $\mathcal{H}_{\mathcal{L}}$ valid for large $\Delta$:
\begin{equation}
	F_{\mathcal{L}}(\Delta) \equiv \int_0^\Delta \text{d}\Delta' \rho_{\mathcal{L}}(\Delta') = \frac{N_0}{2\pi}	\left(\frac{3}{c\Delta}\right)^{1/4} e^{2\pi\sqrt{\frac{c}{3}\Delta}}\big[1 + O(\Delta^{-1/2})\big], \,\,\, \Delta\rightarrow \infty.
\end{equation}

\end{itemize}

\section{Charting Hilbert Space $\mathcal{H}^{\mL}$ associated with invertible TDL $\mathcal{L}$}\label{sec:irrep}
In this section we consider invertible TDLs associated with a global symmetry $G$. In particular, we will be focussing on the case where the symmetry group is finite. The primary goal is to focus on the untwisted sector (we are imposing periodic boundary condition along the spatial circle) estimate the growth of operators which transforms under a particular irreducible representation of the group $G$. Later on we will generalize our result to a given twisted sector, where another TDL is inserted along the temporal direction if the symmetry is non-anomalous.

\subsection{Warm up: $G=\mathbb{Z}_2$}\label{subsec:warm}
The symmetry group $\mathbb{Z}_2$ has two elements: identity $e$ and the element $p$, which squares to Identity. We set up the following notation for any group element $g\in G$: 
\begin{equation}
Z^{\mathcal{L}}(\beta, g)=\text{Tr}\left(ge^{-\beta\left(L_0+\bar{L}_0-\frac{c}{12}\right)}\right)\,.
\end{equation}

Thus for $g=e$ we have the usual partition function while for $g=p$ we have 
\begin{equation}
Z^{\mathcal{L}}(\beta, p)= Z_{\text{even}}(\beta)-Z_{\text{odd}}(\beta)\,,
\end{equation}
where $Z_{\text{even}}$ ($Z_{\text{odd}}$) is the partition function for all the even (odd) operators. Clearly, $Z_{\text{even}}(\beta)+Z_{\text{odd}}(\beta)=Z^{\mathcal{L}}(\beta, e)$. In the usual Cardy formula, we want to have an estimate of partition function at high temperature. Similarly, we want to have an expression for $Z_{\text{even}}(\beta)$ and $Z_{\text{odd}}(\beta)$ in the $\beta\to 0$ limit.\\

We have
\begin{equation}
\begin{aligned}\label{29}
Z_{\text{even}}(\beta)&= \frac{1}{2}\left(Z^{\mathcal{L}}(\beta, e)+Z^{\mathcal{L}}(\beta, p)\right)=\frac{\text{dim}(even)}{|G|}\sum_{g}\chi_{\text{even}}^{*}(g)Z^{\mathcal{L}}(\beta, g)\,,\\ 
 Z_{\text{odd}}(\beta)&= \frac{1}{2}\left(Z^{\mathcal{L}}(\beta, e)-Z^{\mathcal{L}}(\beta, p)\right)=\frac{\text{dim}(odd)}{|G|}\sum_{g}\chi_{\text{odd}}^{*}(g)Z^{\mathcal{L}}(\beta, g)\,.
\end{aligned}
\end{equation}
We remark that $\frac{1}{|G|}\sum_{g}\chi_{\alpha}^{*}(g)Z^{\mathcal{L}}(\beta, g)$ calculates the number (weighted by $e^{-\beta(\Delta-c/12)}$, where $\Delta$ is the conformal weight) of times the irrep $\alpha$ is appearing, and the number of states is obtained by multiplying the dimension of irrep to the quantity. We briefly review the representation theory of finite group in \S\ \ref{srefreview}. For any Abelian group, the dimension of irrep is $1$ always, so it is simpler in that scenario. The reason we wrote it in terms of characters $\chi$ is that they immediately generalize to any finite group. For $\mathbb{Z}_2$ the trivial representation is the one where $\chi_{\text{even}}(g)=1$ for all $g\in G=\mathbb{Z}_2$. The nontrivial irrep is the one where we have $\chi_{\text{odd}}(e)=1$ and $\chi_{\text{odd}}(p)=-1$. For $G=\mathbb{Z}_2$, we have $|G|=2$, the order of the finite group.\\

Before delving into the rigorous Tauberian formalism, let us gain some intuition by doing usual Cardy like analysis. For brevity, let us write $Z_{+}\equiv Z_{\text{even}}$, $Z_-\equiv Z_{\text{odd}}$ and similarly $\chi_{\text{even}}\equiv \chi_{+}$, $\chi_{\text{odd}}\equiv \chi_{-}$; $\text{dim}(even)\equiv d_+$, $\text{dim}(odd)\equiv d_-$. Now in the dual channel, we have
\begin{equation}
\begin{aligned}
Z_{\pm}(\beta\to 0)&=\frac{d_{\pm}}{|G|}\left[\chi^{*}_{\pm}(e)Z_{\mathcal{L}}(\beta'\to\infty, e)+\chi^{*}_{\pm}(p) Z_{\mathcal{L}}(\beta'\to\infty, p)\right]\,, \quad \beta'=\frac{4\pi^2}{\beta}\\
& =\frac{1}{2}\left[Z_{\mathcal{L}}(\beta'\to\infty, e)\pm Z_{\mathcal{L}}(\beta'\to\infty, p)\right]\,, \quad \beta'=\frac{4\pi^2}{\beta}\,.
\end{aligned}
\end{equation}
Here $Z_{\mathcal{L}}(\beta', e)$ is the usual partition function evaluated at the dual temperature $\beta'$. The quantity $Z_{\mathcal{L}}(\beta', p)$ is obtained by doing modular transformation on $Z^{\mathcal{L}}(\beta, p)$. Now $Z^{\mathcal{L}}(\beta, p)$ is not modular invariant, because it has an insertion of TDL along spatial direction. Under $S$ modular transformation, cycles of the torus get exchanged, thus we have a torus configuration where the TDL is along the time direction. We can interpret this as having a defect in the spatial circle. Thus $Z_{\mathcal{L}}(\beta', p)$ is the partition function for the defect Hilbert space.\\

If the ground state in the defect Hilbert space (corresponding to $g\neq e$) has $\Delta>0$, we have
\begin{equation}\label{asymp}
Z_{\pm}(\beta\to 0)=\frac{d_{\pm}}{|G|}\chi^{*}_{\pm}(e)Z_{\mathcal{L}}(\beta'\to\infty, e)=\frac{d_{\pm}^2}{|G|} \exp\left[\frac{\pi^2c}{3\beta}\right]=\frac{1}{2} \exp\left[\frac{\pi^2c}{3\beta}\right]\,.
\end{equation}

Let us pause for a moment and discuss when we can ensure that $\Delta>0$ in the defect Hilbert space. According to \cite{Lin:2019kpn}, if the $\mathbb{Z}_2$ is anomalous, then the spin is constrained to be of the form $\frac{1}{4}+\mathbb{Z}/2$, thus excludes the possibility of having $\Delta=0$ state. Similar argument is true for anomalous $\mathbb{Z}_n$ for any $n$. Since any finite group has a subgroup $\mathbb{Z}_{m}$ for some $m\in\mathbb{Z}_+$, if the subgroup is anomalous, the argument applies and we can not have $\Delta=0$ state in the defect Hilbert space corresponding to that subgroup. If $\mathbb{Z}_2$ is non-anomalous, then we can not apply this argument. Nonetheless, we can gauge the $\mathbb{Z}_2$ group in such scenario to obtain the orbifold theory. We note that $\Delta=0$ states is an even state, so it will be in the even sector of the defect Hilbert space if it is there in defect Hilbert space to begin with. The orbifolded theory has even operators from the usual Hilbert space (untwisted sector) and the even operators from the defect Hilbert space (twisted sector). Now if we assume the uniqueness of the $\Delta=0$ state in the orbifolded theory, the defect Hilbert space can not have any $\Delta=0$ state.The another way to phrase the statement is to demand that the action of symmetry group is faithful, thus the only TDL which commutes with all the operators is the Identity line as explained in the \S\ \ref{review}. In what follows, we will assume this as a generic condition that in the defect Hilbert space $\Delta>0$. We mention that the assumption is true for the Ising model.\\

From \eqref{asymp}, we immediately derive the growth of operators in even and odd sector: 
\begin{align}\label{pointer}
\rho_{\pm}(\Delta) \underset{\Delta\to\infty}{\simeq} \frac{1}{2} \left(\frac{c}{48\Delta^3}\right)^{\frac{1}{4}}\exp\left[2\pi\sqrt{\frac{c\Delta}{3}}\right]\,,
\end{align}
where $\rho_{\pm}$ stands for density of states for even and odd operators respectively.\\

We make some remarks below:\\

$\bigstar$ {\bf Smearing turns $d_{\alpha}$ into $d^2_{\alpha}$}: Presence of symmetry predicts an extra-fold degeneracy of $d_{\alpha}$ where $d_{\alpha}$ is the dimension of $\alpha$ irrep. Thus it is somewhat surprising to find $d_{\alpha}^2$ in the expression for density of states. But as we will show in the next subsection, the expression for the density of states is true only after smearing over an order one window. This smearing\footnote{We thank Raghu Mahajan for discussion along this line.} allows for an effective extra-fold degeneracy of $d_{\alpha}^2$. This becomes particularly clear if we examine the 3-state Potts model ($c=4/5$), which has $S_3$ symmetry (See \cite{Chang:2018iay} for a quick and nice exposition of this theory with an emphasis on TDLs). $S_3$ is a generated by two elements: one element generates the $\mathbb{Z}_3$ symmetry, while the other element acts as $\mathbb{Z}_3$ charge conjugation. There are two doublet of primaries in this CFT sitting in the nontrivial $2$ dimensional $S_3$ representation. Each of the doublet contains a primary of $\mathbb{Z}_3$ charge $\omega$ and a primary of $\mathbb{Z}_3$ charge $\omega^{*}=\omega^2$. One doublet has dimension $2/15$ while the other one has dimension $4/3$. All the descendants of these primaries sit in the same representation. If we consider a window of width $2\delta\gtrapprox1$, it contains descendants of both the doublets. Thus it gives a factor of $2^2=4$. Should we able to resolve the actual density of states, we would have found degeneracy of $2$ as predicted by the actual symmetry. Furthermore, note that for $S_3$, we have $|G|=6$, thus we have a growth of $4/6\rho_0$ for the doublet irrep. From the perspective of $\mathbb{Z}_3$, we are counting all the operators with charge $\omega$ and $\omega^2$, thus we should have a growth of $(1+1)/3\rho_0$, lo and behold $4/6=2/3$. Roughly speaking, the irrep $\alpha$ has to appear $d_{\alpha}$ times in a window of width $2\delta\to1^{+}$, this might hint at some approximate symmetry which emerges only because we smear. The scenario is very much similar to the one present in the calculation of disk partition function of  JT gravity and bulk gauge field theory \cite{Kapec:2019ecr}.
\\

$\bigstar$ It might seem very tempting to discuss the growth of $\rho_+-\rho_-$. Naively, asymptotic growth of $\rho_+-\rho_-$ is controlled by inverse Laplace transformation of $Z^{\mathcal{L}}(\beta\to 0, p)=Z_{\mathcal{L}}(\beta' \to\infty, p)$ \cite{Lin:2019kpn}. Nonetheless this argument does not pass the rigorous treatment of Tauberian since the positivity of $\rho_+-\rho_-$ is not guaranteed, in fact it can in principle widely oscillate. Nonetheless, it is also possible to prove the following as a corollary of the theorem proven in the next section.
\begin{equation}
\begin{aligned}
\bigg| \int_{\Delta-\delta}^{\Delta+\delta}\text{d}\Delta'\ \left[\rho_+(\Delta')-\rho_-(\Delta')\right]\bigg| \leq  \left(\frac{c}{48\Delta^3}\right)^{\frac{1}{4}}\exp\left[2\pi\sqrt{\frac{c\Delta}{3}}\right]
\end{aligned}
\end{equation}
where we have used the extremal functions appearing in \cite{Mukhametzhanov:2020swe} to fix the order one number.\\

$\bigstar$ For $c>1$, the analysis can be made sensitive to Virasoro primaries only. In the following section, we will be generalizing the idea to arbitrary finite group $G$ using the notion of character as well as we will make our analysis rigorous using Tauberian formalism \cite{Baur,Ganguly:2019ksp,Pal:2019zzr}.

\subsection{Arbitrary finite group $G$ ala Tauberian}
\label{subsec:general}
\subsubsection{Untwisted sector}
The partition function for the operators transforming under particular irreducible representation $\alpha$ is given by
\begin{equation}
Z^{\mathcal{L}}_{\alpha}(\beta)=\frac{d_{\alpha}}{|G|}\sum_{g\in G}Z^{\mathcal{L}}(\beta,g)\chi_{\alpha}^{*}(g)\equiv \int_{0}^{\infty}\text{d}\Delta'\ \rho_{\alpha}(\Delta')
\end{equation}
where $d_{\alpha}$ is the dimension of the irrep $\alpha$. Under $S$ modular transformation we have
\begin{equation}\label{def:ch}
Z^{\mathcal{L}}_{\alpha}(\beta) \underset{S}{\to} Z_{\mathcal{L}}{}_{\alpha}(\beta') =\frac{d_{\alpha}}{|G|}\sum_{g\in G}Z_{\mathcal{L}}(\beta',g)\chi_{\alpha}^{*}(g)
\end{equation}
where $\beta'=\frac{4\pi^2}{\beta}$. Our objective is to establish the following theorem:
\begin{theorem} \label{th:irrepcount}
We consider untwisted sector of a CFT admitting a global symmetry under a finite group $G$. The states transforming under the irreducible representation $\alpha$, has an asymptotic growth, which is given by:


\begin{equation}
\begin{aligned}\label{eq:irrepcount}
&\frac{c_- d_{\alpha}^2}{|G|} \rho_0(\Delta)\leq \frac{1}{2\delta}\int_{\Delta-\delta}^{\Delta+\delta}\text{d}\Delta'\ \rho_{\alpha}(\Delta') \leq \frac{c_+d_{\alpha}^2}{|G|}\rho_0(\Delta)
\end{aligned}
\end{equation}
Here $\rho_0(\Delta)$ is defined in \eqref{def:rho0} and $c_{\pm}$ are order one positive numbers.  These numbers can be determined using the extremal functionals appearing in \cite{Mukhametzhanov:2020swe}. In particular, we have $c_{\pm}=1\pm 1/2\delta$. The above statement is true under the assumption that $\mathcal{H}_{\mathcal{L}}(g)$ does not contain $\Delta=0$ state for $g$ not equal to the identity ($e$) element. This ensures that the sum defining $Z_{\mathcal{L}}{}_{\alpha}(\beta')$ in eq.~\eqref{def:ch} is dominated by the $\Delta=0$ state coming from the original Hilbert space, i.e.\!~from $Z_{\mathcal{L}}(\beta',e)$.
\end{theorem}

The proof of the theorem closely resembles the one in the previous section. The leading answer comes from inverse Laplace transformation of $\frac{d_{\alpha}}{|G|}Z_{\mathcal{L}}(\beta',e)\chi_{\alpha}^{*}(e)=\frac{d^2_{\alpha}}{|G|}Z(\beta')$. The only non-trivial part is to show the suppression of the heavy part of $Z_{\mathcal{L}}{}_{\alpha}(4\pi^2/(\beta+\imath t)
) $. Now we have two ingredients, the character and the heavy part of the defect partition function. Like before, the absolute value of the heavy part is dominated by $|t|=\Lambda$. Then we can use the
following chain of inequality 
$$Z_{\mathcal{L}}{}_{H}\left(\frac{4\pi^2\beta}{\beta^2+\Lambda^2},g\right) \leq Z_{\mathcal{L}}\left(\frac{4\pi^2\beta}{\beta^2+\Lambda^2},g\right)= Z^{\mathcal{L}}\left(\frac{\beta^2+\Lambda^2}{\beta},g\right) \leq NZ\left(\frac{\beta^2+\Lambda^2}{\beta}\right)\underset{\beta\to 0}{\simeq}Ne^{\frac{\Lambda^2}{\beta}}\,.$$

On the other hand, the character can be bound using 

\begin{equation}
\begin{aligned}
&|\chi_{\alpha}^{*}(g)|^2 \leq \left(\sum_{g}|\chi_{\alpha}^{*}(g)|^2\right) = |G|\ \quad \Rightarrow\ \quad \frac{1}{G}|\chi_{\alpha}^{*}(g)|\leq \frac{1}{|G|^{1/2}}\,.
\end{aligned}
\end{equation}

Using the above two, we estimate the heavy part integrand for $|t|\leq\Lambda$
\begin{equation}
\begin{aligned}
\bigg|\frac{d_{\alpha}}{|G|}\sum_{g\in G}Z_{\mathcal{L}}{}_{H}(\beta',g)\chi_{\alpha}^{*}(g)\bigg|
&\leq \frac{d_{\alpha}}{|G|^{1/2}} \sum_{g}Z_{\mathcal{L}}{}_{H}\left(\frac{4\pi^2\beta}{\beta^2+\Lambda^2},g\right)\underset{\beta\to 0}{\leq } N|G|^{1/2}d_{\alpha}\exp\left[\frac{\Lambda^2}{\beta}\right] 
\end{aligned}
\end{equation}

Thus we have
\begin{equation}
\bigg|\text{heavy part}\bigg| \leq N|G|^{1/2} d_{\alpha} \exp\left[\frac{\Lambda^2}{\beta}\right]\int_{-\Lambda}^{\Lambda}\text{d}t\ |\hat{\phi}_{\pm}(t)| 
\end{equation}
Again we use the bandlimited functions $\phi_{\pm}$ and choose the support of $\hat{\phi}_{\pm}$ to be $[-\Lambda,\Lambda]$ with $\Lambda=2\pi$. One can choose $\Lambda<2\pi$. In fact by careful treatment, it is possible to choose $\Lambda=2\pi$ and the extremal functions appearing in \cite{Mukhametzhanov:2020swe} to deduce the value of order one numbers $c_{\pm}$ appearing in the theorem.\\

\subsubsection{Twisted sector} 
One can consider the twisted sector by introducing the TDL corresponding to the global symmetry $G$ along temporal direction. This is what we called defect Hilbert space. Now if the symmetry is non-anomalous $G$, we can insert another TDL along the spatial direction and unambiguously resolve the crossing of two TDLs. Within a twisted sector (twisted by a given element $g\in G$) one can estimate the growth of operators transforming under particular irrep of $G$.\\ 

Here we use slightly different notations because now we have to deal insertion of two TDLs. By $Z_{\mathcal{L}}(\beta,g_0,g)$ we mean the partition function evaluated with TDL corresponding to $g_0$ inserted along temporal direction and TDL corresponding to $g$ inserted along spatial direction. We also put in $g_0$ as argument of density of states to remind ourselves that we are dealing with the twisted sector. Thus the partition function $Z^{(\alpha)}_{\mathcal{L}}(\beta,g_0)$ for the operators in the $\alpha$ irrep in the twisted sector is given by
\begin{equation}
Z^{(\alpha)}_{\mathcal{L}}(\beta,g_0)\equiv \frac{d_{\alpha}}{|G|}\sum_{g\in G}Z_{\mathcal{L}}(\beta,g_0,g)\chi_{\alpha}^{*}(g)\equiv \int_{0}^{\infty}\text{d}\Delta'\ \rho_{\alpha}(g_0,\Delta')
\end{equation}
where $d_{\alpha}$ is the dimension of the irrep $\alpha$. Under $S$ modular transformation\footnote{When $g$ and $g_{0}$ are both non-identity elements, under $S$ modular transformation, the relative orientation of the TDLs corresponding to them changes. Hence in the dual channel we have $g_0^{-1}$ inserted along the spatial direction.} we have
\begin{equation}\label{def:chd}
Z^{(\alpha)}_{\mathcal{L}}(\beta,g_0) \underset{S}{\to} \frac{d_{\alpha}}{|G|}\sum_{g\in G}Z_{\mathcal{L}}(\beta',g,g_0^{-1})\chi_{\alpha}^{*}(g)\,,\ \text{where}\ \beta'=\frac{4\pi^2}{\beta}\,.
\end{equation}

The final result is again given by eq.~\eqref{eq:irrepcount}. In particular we have
\begin{theorem}\label{th:irrepcounttw}
We consider twisted sector (twisted by the $g_0\in G$) of a CFT admitting a symmetry (non-anomalous) under a group $G$. 
The \eqref{eq:irrepcount} holds true for the growth of operators in this sector. The assumptions are same as the one in theorem~[\ref{th:irrepcount}].
\end{theorem}

\subsection{TDL associated with continuous symmetry $U(1)$} \label{subsec:u1}
The idea presented above for the finite group can be generalized to continuous group as well. The tricky part is to determine the behavior of $Z^{\mathcal{L}_g}(\beta\to 0,\alpha)=Z_{\mathcal{L}_g}(\beta'\to \infty,\alpha)$. Earlier knowing that for $g\neq e$, the defect Hilbert space has states with $\Delta>0$ only sufficed because we have a sum over group elements. But here we have an integral over the group manifold and as $g\to e$, the ground state of the defect Hilbert space goes to $0$. Thus we need to know the behavior of the ground state of the defect Hilbert space as $g\to e$, to say something concrete. \\

In what follows, we can consider the compact $U(1)$ group, which is generated by $J\equiv J_0-\bar{J}_0$, coming from the Kac-Moody algebra. For a nice discussion related to compact vs non compact we refer the readers to \cite{Benjamin:2020swg}. The partition function is given by
\begin{equation}
Z^{\mathcal{L}_g}(\beta,\nu,\bar{\nu})=\sum_{n,J}e^{-\beta \left(\Delta_n-\frac{c}{12}\right)} e^{2\pi\imath\nu J}
\end{equation}
where $g=e^{2\pi\imath\nu}\in G=U(1)$ and $\nu\in(-1/2,1/2]$.  Usually we think of this as partition function for Grand Canonical ensemble. Alternatively, we can think of this as a partition function evaluated on Torus with insertion of TDLs corresponding to $U(1)$.

We wish to estimate asymptotic growth of states with a definite $J$. We write down a partition function for a fixed $J\equiv  Q$:
\begin{equation}
Z^{\mathcal{L}_g}_{Q}(\beta)= \int_{-1/2}^{1/2}\text{d}\nu\ e^{-2\pi\imath\nu Q}Z^{\mathcal{L}_g}(\tau)=\sum_{n} d_{n,Q} e^{-\beta \left(\Delta_n-\frac{c}{12}\right)}
\end{equation}
We pause here to comment about the integral range of $\nu$, i.e. $\nu \in(-1/2,1/2]$. This implies that we are considering ``single'' cover of $U(1)$ and all the charges are integer. We further assume that this action is faithful. Thus we exclude scenarios like where all the charges are even. Instead of ``single'' cover,  we can also consider $N\in\mathbb{Z}_+$ cover , so that possible charges are of the form $\frac{q}{N}$ with $N-1>|q|\in\mathbb{N}$; in that scenario the $\nu$ integral would have been from $-N/2$ to $N/2$ with a multiplicative factor of $\frac{1}{N}$ for correct normalization. In this way of thinking, the scenario, where all the charges are even can be treated as effectively making $N=\frac{1}{2}$. In what follows, we will be considering $N=1$ case. Without loss of generality, we also assume the spectra is invariant under $Q\to-Q$ as they correspond to taking $\nu\to-\nu$. As an example, readers might keep in mind compact boson with level $k=1$ and radius $R=2$, where the charge under $J_0$ is $\frac{e}{R}+\frac{mR}{2}$ and the charge under $\bar{J}_0$ is $\frac{e}{R}-\frac{mR}{2}$ with $e,m \in\mathbb{Z}$.\\

Modular transformation of $Z^{\mathcal{L}_g}(\beta)$ gives us the partition function of the defect Hilbert space and we have
\begin{equation}
\begin{aligned}
Z_{\mathcal{L}_g}{}_{Q}\left(\frac{4\pi^2}{\beta}\right)&=\int_{-1/2}^{1/2}\text{d}\nu\ e^{-2\pi\imath\nu Q}Z_{\mathcal{L}_g}\left(\frac{4\pi^2}{\beta}\right)\\
&=\sum_{n,q,\bar{q}} d_{n,q,\bar{q}} \int_{-1/2}^{1/2}\text{d}\nu\ e^{-2\pi\imath \nu Q} e^{-\frac{4\pi^2}{\beta}\left(\Delta_n-\frac{c}{12}+ k\nu^2 -\nu( q+\bar{q})\right)}\,, 
\end{aligned}
\end{equation}
where $k$ is a parameter coming from the Kac-Moody algebra, $q$,$\bar{q}$ are the charge under $J_0$ and $\bar{J_0}$. We want to evaluate this integral in the $\beta\to0$ limit. 

\begin{equation}
\begin{aligned}
Z_{\mathcal{L}_g}{}_{Q}\left(\frac{4\pi^2}{\beta}\right)\simeq\sqrt{\frac{\beta}{4\pi k}}\sum_{\underset{|q+\bar{q}|\leq' k}{\widehat{\Delta}_n}} d_{\widehat{\Delta}_n,q,\bar{q}}\ e^{-\frac{4\pi^2}{\beta}\left(\widehat{\Delta}_n-\frac{c}{12}\right)-\beta\frac{Q^2}{4k}-\imath\pi\frac{Q(q+\bar{q})}{k}} 
\end{aligned}
\end{equation}
where $\widehat{\Delta}_n=\left(\Delta_n-\frac{(q+\bar{q})^2}{4k}\right)$. The prime over $\leq$ indicates whenever $q+\bar{q}$ becomes $\pm k$, there is a factor of $\frac{1}{2}$ associated with the $\nu$ integral. The crucial point is to observe that the states in the defect Hilbert space has dimension $\widehat{\Delta}+(\nu-\frac{q+\bar{q}}{2k})^2$. Thus in the $\beta\to 0$ limit, the $\nu$ integral can contribute only if $(\nu-\frac{q+\bar{q}}{2k})=0$ for some $\nu \in (-1/2,1/2]$. Thus the sum over $q,\bar{q}$ is restricted. Thus the leading piece is given by $\widehat{\Delta}=0$ states. Of course $\Delta=q=\bar{q}=0$ would contribute. We observe that the unitarity bound tells us that
\begin{equation}
\Delta \geq \frac{q^2+\bar{q}^2}{2k} \geq \frac{(q+\bar{q})^2}{4k}\,,
\end{equation}
where the saturation of the second inequality can happen only if $q=\bar{q}$. Thus the states that would contribute to the leading order is given by $\widehat{\Delta}=0, q=\bar{q}, |q|\leq' \frac{k}{4}$. Hence we have 
\begin{equation}
Z^{\mathcal{L}_g}_{Q,\bar{Q}}(\beta\to 0) \simeq N_0\sqrt{\frac{\beta}{4\pi k}} e^{\frac{\pi^2c}{3\beta}}
\end{equation}
where $N_0=\sum_{q=\bar{q},|q|\leq' k/2} w(q)e^{- \frac{2\pi\imath qQ}{k}}$ and $w(q)=1$ if $|q|< k/2$ and $w(q)=\frac{1}{2}$ if $|q|=\frac{k}{2}$. Since we have assumed $q\to-q$ symmetry, $N_0$ is a real number. $N_0=1$ if only such state is the vacuum. In what follows, we will assume that this is the case\footnote{ We note that if the action of $U(1)$ is not faithful, for example, if all the charges are even, then for odd charges, the asymptotic expression should give $0$, as a result $N_0$ should have been equal to $0$, in those cases the phases in the sum defining $N_0$ play a key role. } and $N_0=1$. Strictly speaking, in the Tauberian analysis, we would require the above argument to hold for complex $\beta+\imath t$.\\

The next step is to split up the Hilbert space into the light and the heavy sector. Now we divide the Hilbert space using the quantity $\widehat{\Delta}\equiv \Delta-\frac{q^2+\bar{q}^2}{4k}$. The light sector is defined as $\widehat{\Delta}+(\nu-\frac{q}{2k})^2\leq c/12$ while the heavy sector is the one with $\widehat{\Delta}+(\nu-\frac{q}{2k})^2>c/12$. Thereafter, we restrict our attention to the heavy sector and show it is suppressed. Recall the quantity related to the heavy sector that appears in the Tauberian analysis is following:
$$I\equiv e^{\beta(\Delta\pm\delta-c/12)}\bigg|\int_{-\Lambda}^{\Lambda}\text{d}t\ e^{-\imath tc/12}\hat{\phi}(t)\ \sum_{n,q,\bar{q}} d_{n,q,\bar{q}}  \int_{-1/2}^{1/2}\text{d}\nu\ e^{-2\pi\imath \nu Q} e^{-\frac{4\pi^2}{\beta+\imath t}\left(\Delta_n-\frac{c}{12}+ k \nu^2 + \nu(q+\bar{q})\right)}\bigg|\,.$$

Now we pull in the absolute value inside the integral and notice the exponential is maximized for $|t|=\Lambda$. Thus we have
\begin{equation}
\begin{aligned}
I&\leq e^{\beta(\Delta\pm\delta-c/12)}\int_{-1/2}^{1/2}\text{d}\nu\ \sum_{\underset{\text{heavy}}{n,q,\bar{q}}} d_{n,q,\bar{q}}\ e^{-\frac{4\pi^2\beta}{\beta^2+\Lambda^2}\left(\Delta_n-\frac{c}{12}+ k\nu^2 + \nu(q+\bar{q})\right)}\int_{-\Lambda}^{\Lambda}\text{d}t\ |\hat{\phi}(t)| \\
&\leq e^{\beta(\Delta\pm\delta-c/12)}\int_{-1/2}^{1/2}\text{d}\nu\ Z_{\mathcal{L}_{g}}\left(\frac{4\pi^2\beta}{\beta^2+\Lambda^2}\right)\int_{-\Lambda}^{\Lambda}\text{d}t\ |\hat{\phi}(t)|\\
&=e^{\beta(\Delta\pm\delta-c/12)}\int_{-1/2}^{1/2}\text{d}\nu\ Z^{\mathcal{L}_{g}}\left(\frac{\beta^2+\Lambda^2}{\beta},\nu\right)\int_{-\Lambda}^{\Lambda}\text{d}t\ |\hat{\phi}(t)|\\
&\leq e^{\beta(\Delta\pm\delta-c/12)}\int_{-1/2}^{1/2}\text{d}\nu\ Z^{\mathcal{L}_{g}}\left(\frac{\beta^2+\Lambda^2}{\beta},\nu=0\right)\int_{-\Lambda}^{\Lambda}\text{d}t\ |\hat{\phi}(t)|\\
&\underset{\beta\to0}{\simeq}  e^{\beta(\Delta\pm\delta-c/12)} e^{\frac{\pi^2c}{3\beta}\frac{\Lambda^2}{4\pi^2}}\int_{-\Lambda}^{\Lambda}\text{d}t\ |\hat{\phi}(t)| \underset{\beta=\pi\sqrt{\frac{c}{3\Delta}}}{\simeq} \exp\left[\frac{\Lambda^2}{4\pi^2} 2\pi \sqrt{\frac{c\Delta}{3}}\right]
\end{aligned}
\end{equation}
We will see that the suppression requires $\Lambda < 2\pi$. The light sector produces the leading Cardy like behavior for density of states $\rho_{Q,\bar{Q}}(\Delta')$ of operators with fixed order one charge $Q,\bar{Q}$ and large conformal dimension $\Delta$. This can be obtained by doing the following integral and realizing that the integral is dominated by $t=0$ in the $\beta\to 0$ limit:
\begin{equation}
\begin{aligned}
&e^{\beta(\Delta\pm\delta-c/12)}\int_{-\Lambda}^{\Lambda} e^{-\imath t c/12} \sqrt{\frac{\beta+\imath t}{4\pi k}} \exp\left[\frac{\pi^2c}{3(\beta+\imath t)}\right] \hat{\phi}_{\pm}(t)\\
&=e^{\beta(\Delta\pm\delta-c/12)}\hat{\phi}_{\pm}(0)\exp\left[\frac{\pi^2c}{3\beta}\right]  \sqrt{\frac{\beta}{4\pi k}} \left(\frac{3}{\pi c}\right)^{1/2}\beta^{3/2}\\
&=2\pi\hat{\phi}(0) \sqrt{\frac{1}{k}}\left(\frac{1}{4\Delta}\sqrt{\frac{c}{3}}\exp\left[2\pi\sqrt{\frac{c\Delta}{3}}\right]\right)
\end{aligned}
\end{equation}
where the factor $(\frac{3}{\pi c})^{1/2}\beta^{3/2}$ comes from the integrating over the fluctuation around the saddle at $t=0$.
Thus we have following estimate:
\begin{equation}
\begin{aligned}\label{eq:u1count}
&c_-\sqrt{\frac{c}{48k\Delta^2}}\exp\left[2\pi\sqrt{\frac{c\Delta}{3}}\right]\leq\frac{1}{2\delta}\int_{\Delta-\delta}^{\Delta+\delta}\text{d}\Delta'\ \rho_{Q}(\Delta')\leq c_+\sqrt{\frac{c}{48k\Delta^2}}\exp\left[2\pi\sqrt{\frac{c\Delta}{3}}\right]\,.
\end{aligned}
\end{equation}
Here $c_{\pm}=\frac{2\pi}{2\delta}\hat{\phi}_{\pm}(0)$ is order one positive number. \\

We conclude this section with two final remarks that one can generalize the analysis for Virasoro primaries for CFT with $c>1$ and one can generalize this to large central charge.

\section{Tauberian for Vector-valued modular function}\label{sec:vvm}
The results for the finite group can nicely be encapsulated in terms of something known as vector valued modular function. The vector-valued modular function $\mathbf{Z}$ obeys the following transformation law under $S$ modular transformation:
\begin{equation}\label{vvm}
\mathbf{Z}\left(\frac{4\pi^2}{\beta+\imath t}\right)=\mathbf{F}\cdot\mathbf{Z}(\beta+\imath t)
\end{equation}
where $\mathbf{Z}$ is a column vector consisting of bunch of functions and $\mathbf{F}$ is a constant ($\beta$ independent) matrix. The condition $S^2=\mathbb{I}$ boils down to $\mathbf{F}^2=\mathbf{I}$.\\

$\bullet$ In the example of CFT with $\mathbb{Z}_2$ symmetry we can consider $$\mathbf{Z}=(Z_+,Z_-,Z_{p})^{T}\,,$$ where $Z_{\pm}$ are the partition functions for even and odd operator and $Z_{p}$ is the defect Hilbert space with insertion of non-identity $\mathbb{Z}_2$ TDL. The matrix $\mathbf{F}$, in this case, is given by
\begin{equation}
\mathbf{F}=\frac{1}{2}\begin{pmatrix*}
1&\phantom{-}1 &\phantom{-}1 \\
1&\phantom{-}1 &-1 \\
2&-2 &\phantom{-}0
\end{pmatrix*}
\end{equation}
and $\mathbf{F}^2$ is indeed identity.

$\bullet$ For a generic compact group $G$, the vector $\mathbf{Z}$ will have $2k-1$ entry , where $k$ is the number of conjugacy classes of the group $G$. The $k$ entries correspond to $k$ different irreps (recall the number of conjugacy class is equal to number of irreps) and $k-1$ entries correspond to the partition function for the defect Hilbert space with insertion of non-identity element. It suffices to consider one representative element from each conjugacy class as the partition function with insertion of TDL along spatial direction is sensitive to conjugacy class only. For $\mathbb{Z}_{n}$, we have $n$ different conjugacy classes, i.e. $n$ irreps.\\

To estimate the growth of operators in each of the sectors, we define a vector valued density of states $\vec{\rho}(\Delta)$. The upper bound (the lower bound is similar) on the integral of $\vec{\rho}(\Delta)$ is given by a matrix inequality 
\begin{equation}
\int_{\Delta-\delta}^{\Delta+\delta}\text{d}\Delta'\ \vec{\rho}(\Delta') \leq e^{\beta(\Delta+\delta)}\int_{-\Lambda}^{\Lambda}\text{d}t\ \mathbf{F}\cdot\left[\mathbf{Z}\left(\frac{4\pi^2}{\beta+\imath t}\right)\right] e^{-\imath t c/12}\phi_{+}(t)
\end{equation}

Thus we need to estimate the integrals of the form $$e^{\beta(\Delta+\delta)}\int_{-\Lambda}^{\Lambda}\text{d}t\ \mathbf{F}\cdot\left[\mathbf{Z}\left(\frac{4\pi^2}{\beta+\imath t}\right)\right] e^{-\imath t c/12}\phi_{+}(t)$$
in the $\beta\to 0$ limit. 

At this point, we separate out the light contribution and the heavy contribution in the usual way. If we further assume that $\Delta=0$ state appears in one and only one of the sectors, without loss of generality we can keep it as first entry. Then the light sector $\mathbf{Z}^{(L)}$ will give 

$$\int_{\Delta-\delta}^{\Delta+\delta}\text{d}\Delta'\ \vec{\rho}(\Delta') \leq e^{\beta(\Delta+\delta)}\int_{-\Lambda}^{\Lambda}\text{d}t\ \mathbf{F}\cdot\left[\mathbf{Z}^{(L)}\left(\frac{4\pi^2}{\beta+\imath t}\right)\right] e^{-\imath t c/12}\phi_{+}(t)=2\pi\hat{\phi}_{+}(0)\rho_0(\Delta)\vec{\mathbf{F}}_{1}\,,$$
where $\vec{\mathbf{F}}_{1}$ is the first column of the $\mathbf{F}$ matrix. This determines the parameter $N_0$ (or $d^2_{\alpha}$) appearing previously.\\

We still need to show that the heavy sector $\mathbf{Z}^{(H)}$ gives a suppressed contribution in magnitude. In order to achieve that we will use that $|\mathbf{F}_{ij}|<K_{i}$. This is true for all the calculations done previously and generically true because $\mathbf{F}$ is finite matrix and $\mathbf{F}^2=\mathbf{I}$. A more mathematical way to saying this is that $$||\mathbf{F}||_{\infty}=\text{Max}_{i}\left\{\sum_{j}|\mathbf{F}_{ij}|\right\}\ \text{is finite}\,.$$ We note that  for $|t|\leq\Lambda$,
\begin{equation}
\bigg|\left(\mathbf{F}\cdot\left[\mathbf{Z}^{(H)}\left(\frac{4\pi^2}{\beta+\imath t}\right)\right]\right)_{i}\bigg| \leq \sum_{j}|\mathbf{F}_{ij}|\bigg|\mathbf{Z}^{(H)}_{j}\left(\frac{4\pi^2}{\beta+\imath t}\right)\bigg| \leq ||\mathbf{F}||_{\infty}\sum_{j}\mathbf{Z}^{(H)}_{j}\left(\frac{4\pi^2\beta}{\beta^2+\Lambda^2}\right) \,.
\end{equation}

To estimate the sum appearing in the rightmost, we observe that 
\begin{equation}
\sum_{j}\mathbf{Z}^{(H)}_{j}\left(\frac{4\pi^2\beta}{\beta^2+\Lambda^2}\right) \leq \sum_{j}\mathbf{Z}_{j}\left(\frac{4\pi^2\beta}{\beta^2+\Lambda^2}\right) = \sum_{j} \mathbf{F}\cdot \left[\mathbf{Z}\left(\frac{\Lambda^2+\beta^2}{\beta}\right) \right]\,.
\end{equation}

Again we use the fact that $\Delta=0$ appears in one and only one sector to have
$$\sum_{j} \mathbf{F}\cdot \left[\mathbf{Z}\left(\frac{\Lambda^2+\beta^2}{\beta}\right) \right]\simeq \mathbf{Z}_{1}\left(\frac{\Lambda^2+\beta^2}{\beta}\right) \sum_{j}\mathbf{F}_{j1}\simeq e^{\frac{\Lambda^2c}{12\beta}}\,.$$

Choosing $\Lambda<2\pi$ suppress the heavy part. Thus we arrive at our general theorem.

\begin{theorem}\label{th:vvm}
We consider vector valued modular function as defined in \eqref{vvm}. Each entry in the column vector $\mathbf{Z}$ denotes different sector of the CFT. The growth of operators in each of these sectors obey the following inequality:
\begin{equation}\label{eq:vvm}
c_-\rho_0(\Delta)\mathbf{F}_{j1}\leq \frac{1}{2\delta}\int_{\Delta-\delta}^{\Delta+\delta}\text{d}\Delta'\ \rho_{j}(\Delta')\leq c_+\rho_0(\Delta)\mathbf{F}_{j1}\,,
\end{equation}
where $c_{\pm}$ are order one numbers. These numbers can be determined using the extremal functionals appearing in \cite{Mukhametzhanov:2020swe}. In particular, we have $c_{\pm}=1\pm 1/2\delta$.
\end{theorem}

One can further apply similar technique to any rational CFT where characters are indeed vector-valued modular functions, this would facilitate estimation of growth of descendants for each primary (primary of the full chiral algebra). For $c<1$, one can apply this to Minimal models and estimate the growth of descendants of each Virasoro primary.

\section*{Acknowledgements}
The authors acknowledge Ken Intriligator and John McGreevy, Shu-Heng Shao for fruitful discussions and encouragement. SP thanks Daniel Kapec, Raghu Mahajan for discussion related to JT gravity with bulk gauge field and Baur Mukhametzhanov for numerous stimulating generic discussions related to Tauberian theorems. ZS acknowledges the support from the US Department of Energy (DOE) under cooperative research agreement DE-SC0009919 and Simons Foundation award  \#568420. SP acknowledges the support from Ambrose Monell Foundation and DOE grant DE-SC0009988.

{\appendix
\section{Verification}\label{sec:verify}
\subsection{ Ising CFT with $\mathbb{Z}_2$ symmetry}
The Ising model has three primaries: $\mathbb{I}, \epsilon, \sigma.$ Under the $\mathbb{Z}_2$ symmetry, $\mathbb{I}$ (with $h=\bar{h}=0$) and $\epsilon$ (with $h=\bar{h}=\frac{1}{2}$) are even while the $\sigma$ (with $h=\bar{h}=\frac{1}{16}$) is odd. The central charge is given by $c=\frac{1}{2}$. Let us denote the characters corresponding to $\mathbb{I}, \epsilon, \sigma$ as $\chi_0,\chi_{1/2},\chi_{1/16}$ respectively. A nice exposition of the Ising model in context of TDLs can be found in \cite{Lin:2019kpn}. Here we briefly recapitulate the necessary ingredients for verifying our formulas \eqref{eq:irrepcount} and \eqref{eq:defectcount} against the Ising model.\\

The partition function of the Ising model with TDL (name it $\eta$) corresponding to $\mathbb{Z}_2$ group element (the identity element $e$ and the non-identity element $p$) inserted along the spatial direction is given by 
\begin{equation}
\begin{aligned}
Z^{\eta}(\beta,e)&=Z(\beta)=|\chi_{0}|^2+|\chi_{1/2}|^2+|\chi_{1/16}|^2\,,\\
Z^{\eta}(\beta,p)&=|\chi_{0}|^2+|\chi_{1/2}|^2-|\chi_{1/16}|^2\,.
\end{aligned}
\end{equation}

$\bullet$ Irreps: The growth of the even and the odd operators (denoted by $\rho_{\pm}$ respectively) are controlled by $\frac{1}{2}\left(Z^{\eta}(\beta,e)\pm Z^{\eta}(\beta,p)\right)$ in the $\beta\to 0$ limit.
From eq.~\eqref{eq:irrepcount}, we have
\begin{equation}
s_{-}(\delta)\leq \log\left[\frac{1}{2\delta}\int_{\Delta-\delta}^{\Delta+\delta}\text{d}\Delta' \rho_{\pm}(\Delta') \right]-2\pi\sqrt{\frac{\Delta}{6}}-\frac{1}{4}\log\left(\frac{1}{96\Delta^3}\right)+\log(2) \leq s_{+}(\delta)
\end{equation}
where $s_{\pm}=\log(c_\pm)$. We use the value of $c_{\pm}$ presented in \cite{Ganguly:2019ksp}. We verify the above inequality in fig.~\ref{Growth of Odd and Even operators}.
\begin{figure}[H]
\centering
\includegraphics[scale = 0.4]{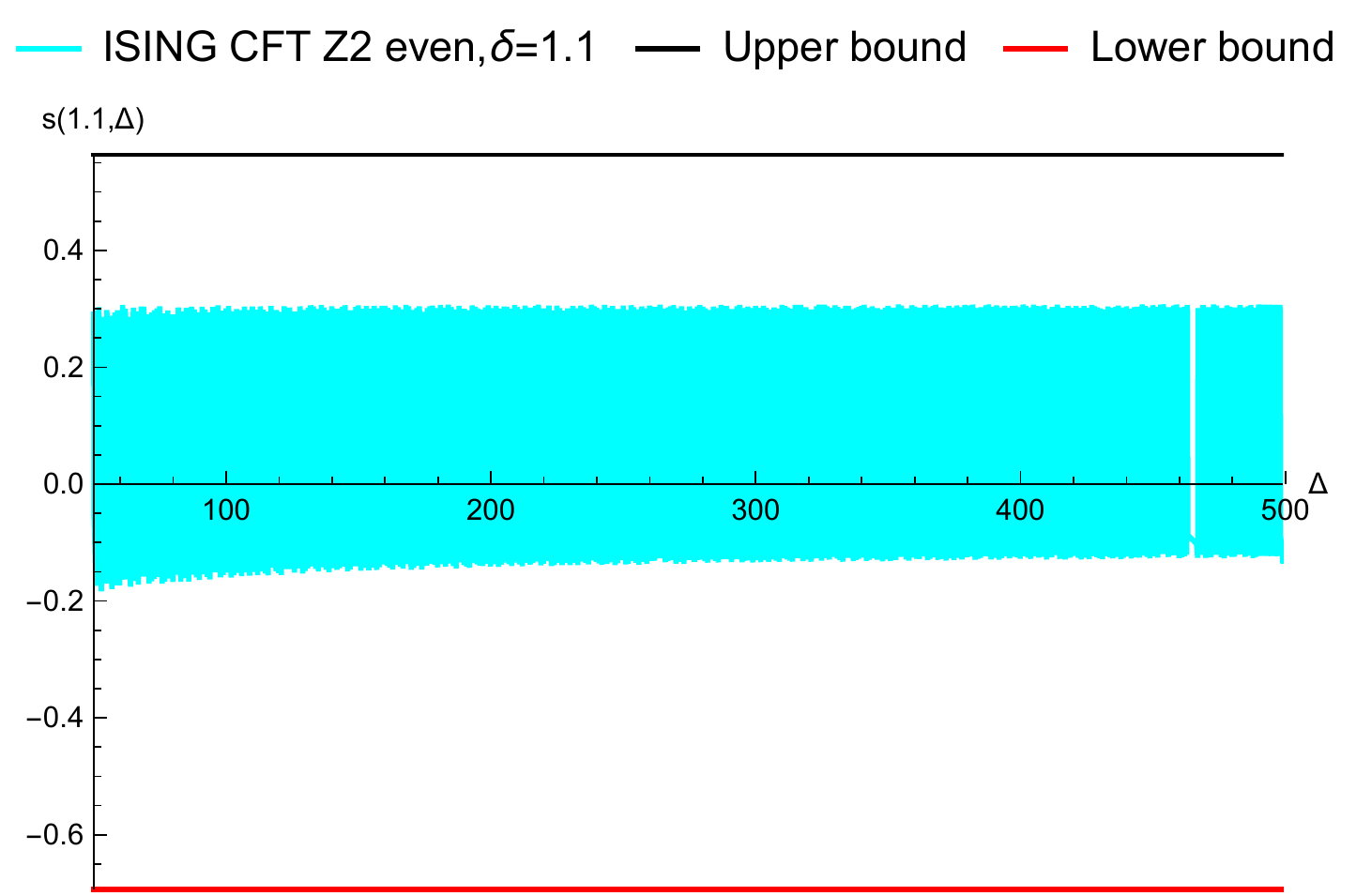}
\includegraphics[scale = 0.4]{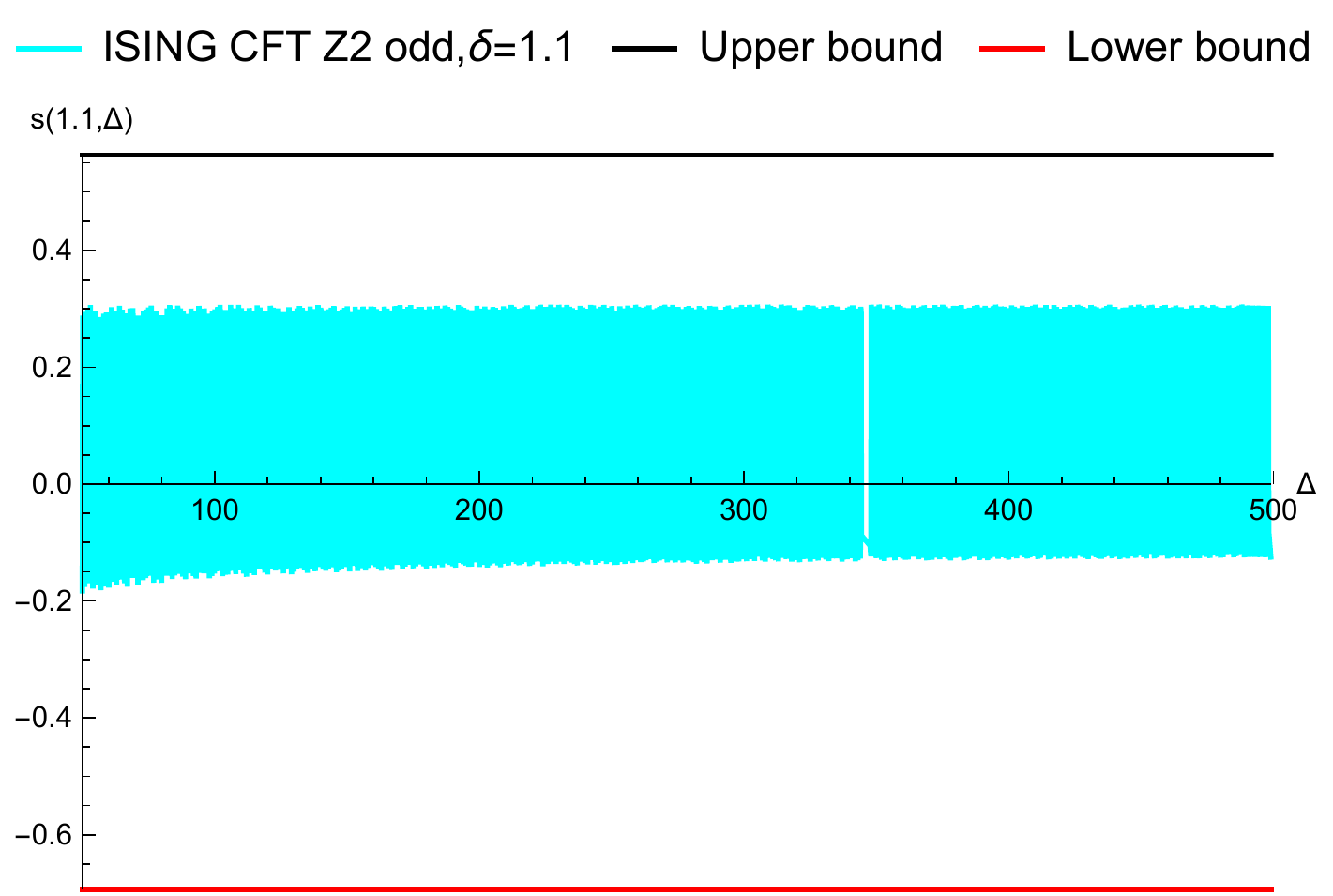}
\caption{Here, we estimate the number of even and odd operators (under $\mathbb{Z}_2$) in Ising CFT. We plot the logarithm of the ratio of actual number of operators in the interval of size $2\delta=2.2$ and the leading prediction from Tauberian-Cardy analysis (in cyan). We see that they are well within the bounds (the black and the red line) as predicted in the main text.}
\label{Growth of Odd and Even operators}
\end{figure}

$\bullet$ Defect Hilbert space: The partition function corresponding to the defect Hilbert space is given by the $S$ modular transformation of $Z^{\eta}(\beta,p)$:
\begin{equation}
Z_{\eta}(\beta)=\chi_0\bar{\chi}_{1/2}+\chi_{1/2}\bar{\chi}_0+|\chi_{1/16}|^2\,.
\end{equation}
The Virasoro primaries have weights $(0,1/2)$, $(1/2,0)$ and $(1/16,1/16)$. We note that there is no $\Delta=0$ state in the defect Hilbert space. The states with $\Delta=1/2$ corresponds to Fermions. We can verify following estimate of the growth of number of operators in the defect Hilbert space (defect corresponding to $\mathbb{Z}_2$, here the TDL is extended along the time direction) of the Ising CFT:
\begin{equation}
s_{-}(\delta)\leq \log\left[\frac{1}{2\delta}\int_{\Delta-\delta}^{\Delta+\delta}\text{d}\Delta' \rho_{\mathcal{H}_{\eta}}(\Delta') \right]-2\pi\sqrt{\frac{\Delta}{6}}-\frac{1}{4}\log\left(\frac{1}{96\Delta^3}\right) \leq s_{+}(\delta)
\end{equation}
where $s_{\pm}=\log(c_\pm)$. The above follows from eq.~\eqref{eq:defectcount}. Again we use the value of $c_{\pm}$ presented in \cite{Ganguly:2019ksp} and verify the inequality in fig.~\ref{defectcount}.
\begin{figure}[H]
\centering
\includegraphics[scale = 0.5]{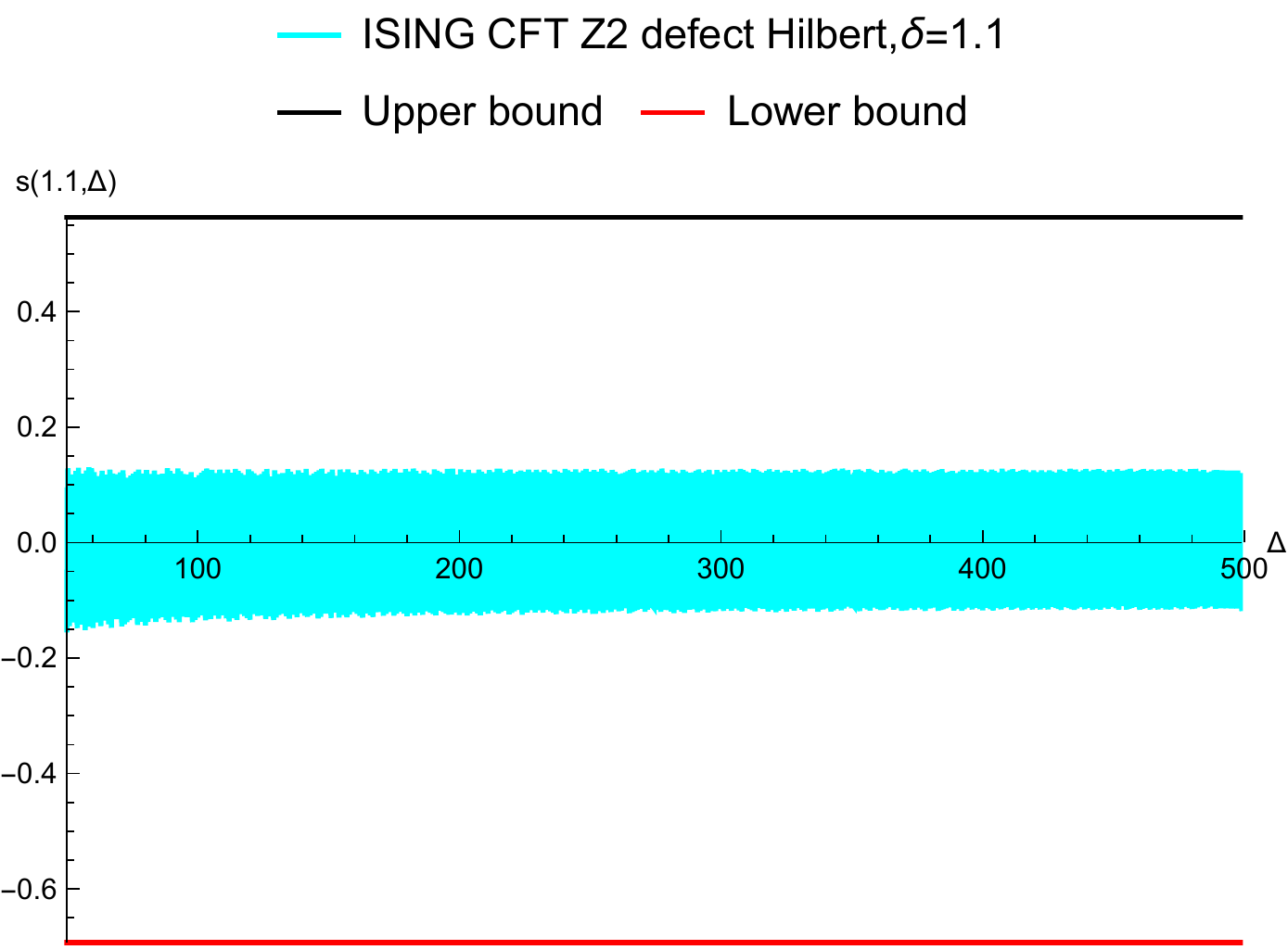}
\caption{Here, we estimate the number of operators in the defect Hilbert space corresponding to  $\mathbb{Z}_2$ in Ising CFT. We plot the logarithm of the ratio of actual number of operators in the interval of size $2\delta=2.2$ and the leading prediction from Tauberian-Cardy analysis (in cyan). We see that they are well within the bounds (the black and the red line) as predicted in the main text.}
\label{defectcount}
\end{figure}

The Ising model also has a duality defect line $\widehat{N}$. This is non invertible TDL. The fusion rule is given by $\widehat{N} \times\widehat{N}=\mathbb{I}+\eta$, thus the action of $\widehat{N}$ is given by

$$\widehat{N}|\text{even}\rangle= \sqrt{2}|\text{even}\rangle\,,\quad \widehat{N}|\text{odd}\rangle= 0\rangle\,.$$

The growth of the operators in the defect Hilbert space corresponding to the duality line can be estimated via eq.~\eqref{eq:defectcount}:
\begin{equation}
s_{-}(\delta)\leq \log\left[\frac{1}{2\delta}\int_{\Delta-\delta}^{\Delta+\delta}\text{d}\Delta' \rho_{\mathcal{H}_{\widehat{N}}}(\Delta') \right]-2\pi\sqrt{\frac{\Delta}{6}}-\frac{1}{4}\log\left(\frac{1}{96\Delta^3}\right)-\frac{1}{2}\log(2) \leq s_{+}(\delta)\,,
\end{equation}
which we verify in the fig.~\ref{defectcountduality}.
\begin{figure}[H]
\centering
\includegraphics[scale = 0.6]{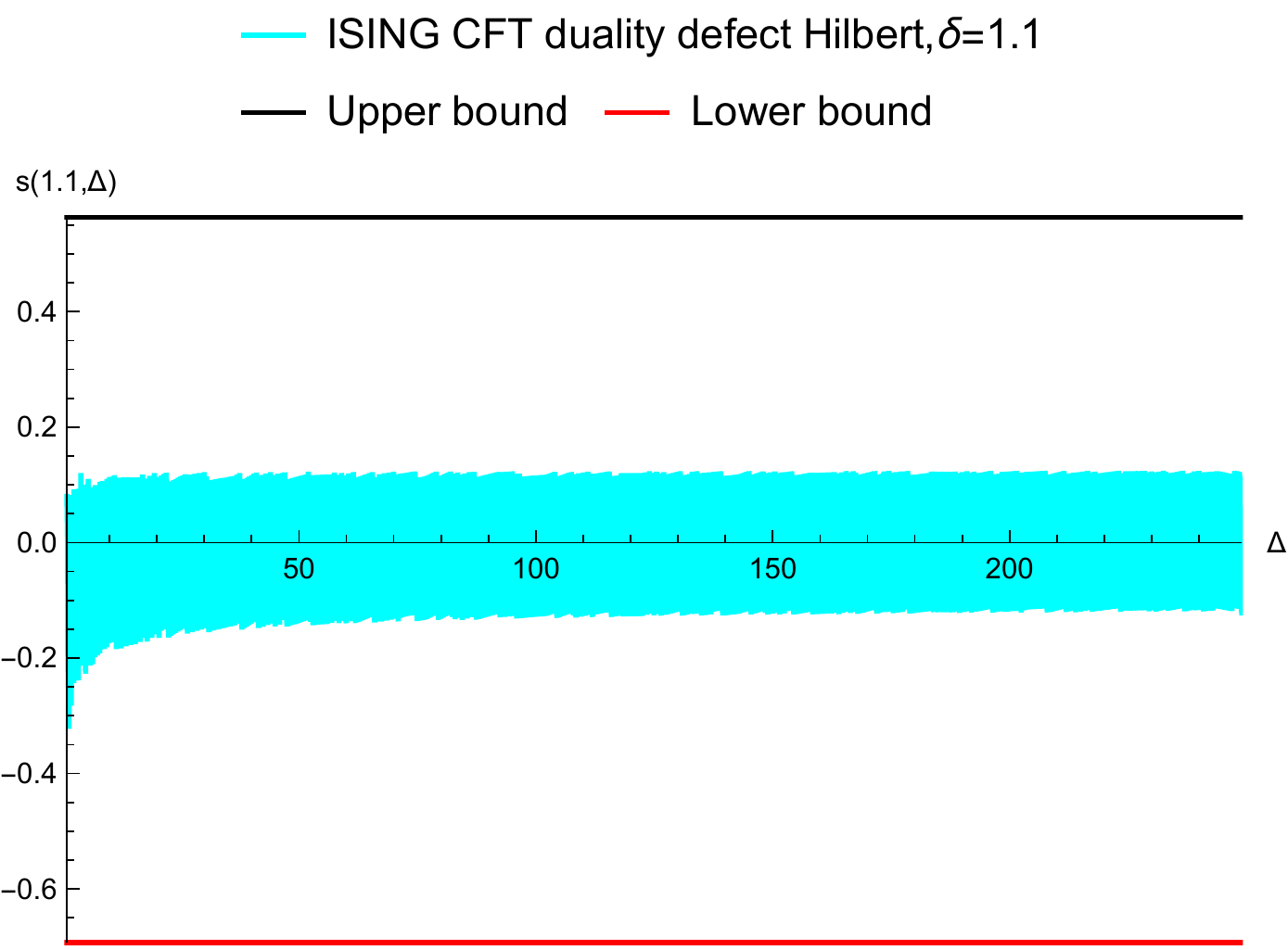}
\caption{Here, we estimate the number of operators in the defect Hilbert space corresponding to the duality defect line $\widehat{N}$ in the Ising CFT. We plot the logarithm of the ratio of actual number of operators in the interval of size $2\delta=2.2$ and the leading prediction from Tauberian-Cardy analysis (in cyan). We see that they are well within the bounds (the black and the red line) as predicted in the main text.}
\label{defectcountduality}
\end{figure}

\subsection{Compact Boson at $R=\frac{1}{2}$ with $U(1)$ symmetry}
For compact Boson at radius $R=\frac{1}{2}$, the $U(1)$ generated by $J_0-\bar{J}_0$ acts faithfully. The partition function for the charge $Q$ is given by 
\begin{align}
Z^{Q}(q)=q^{\frac{m^2}{4}-\frac{1}{12}}\left[\frac{\theta_{3}(q)}{\eta^2}\right]=q^{\frac{m^2}{4}-\frac{1}{12}}\left(1+4q+9q^2+20q^3+O(q^4)\right)\,.
\end{align}
For compact boson $k=1$, thus the growth of operators with charge $Q$ is given by 
\begin{equation}
s_{-}(\delta)\leq \log\left[\frac{1}{2\delta}\int_{\Delta-\delta}^{\Delta+\delta}\text{d}\Delta' \rho_{\mathcal{H}_{\eta}}(\Delta') \right]-2\pi\sqrt{\frac{\Delta}{3}}-\log\left(\frac{1}{4\Delta}\right)+\frac{1}{2}\log(3k) \leq s_{+}(\delta)\,,
\end{equation} 
which follows from eq.~\eqref{eq:u1count}. This is verified in fig.~\ref{cb}.
\begin{figure}[H]
\centering
\includegraphics[scale = 0.5]{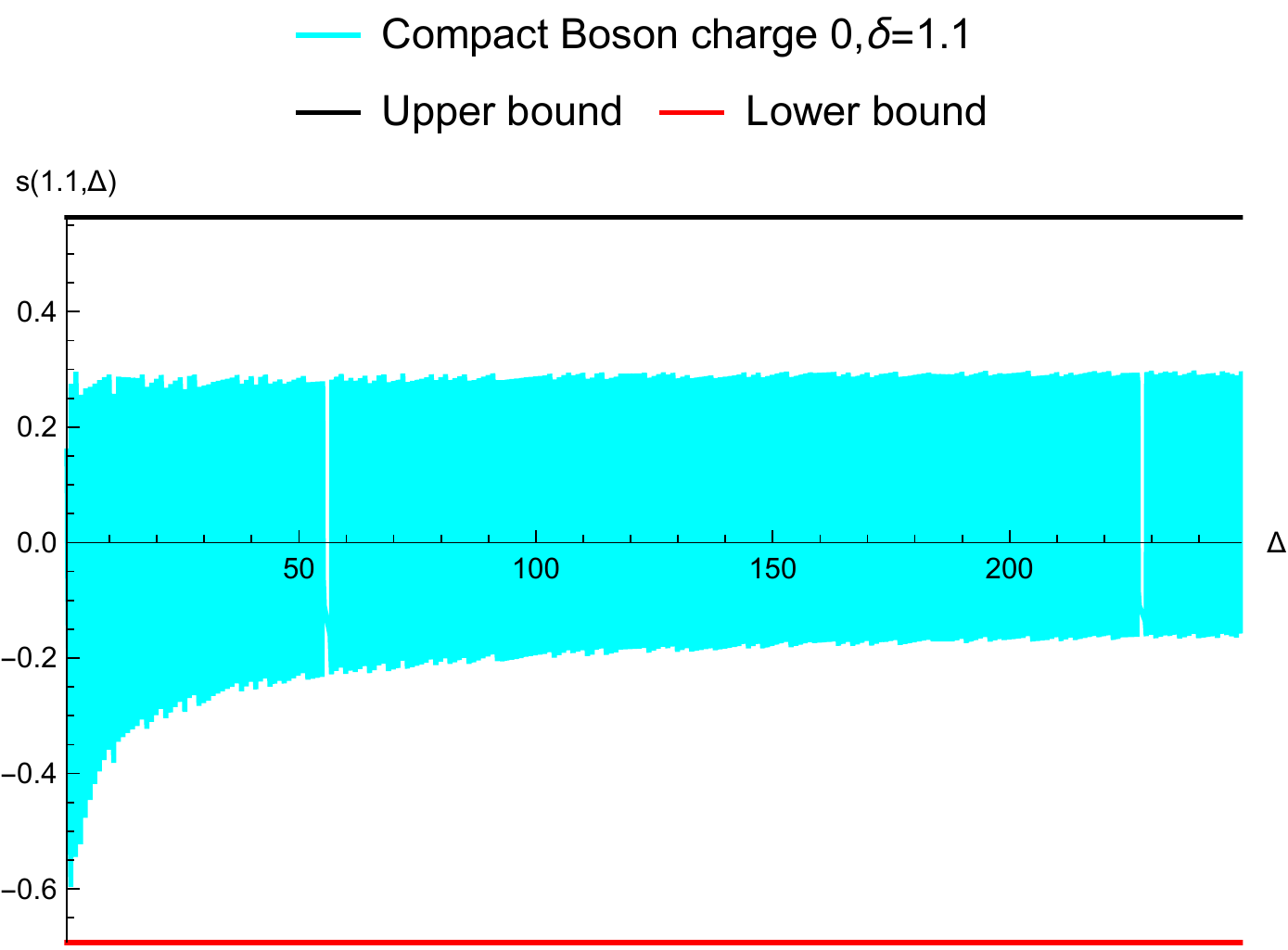}
\includegraphics[scale = 0.5]{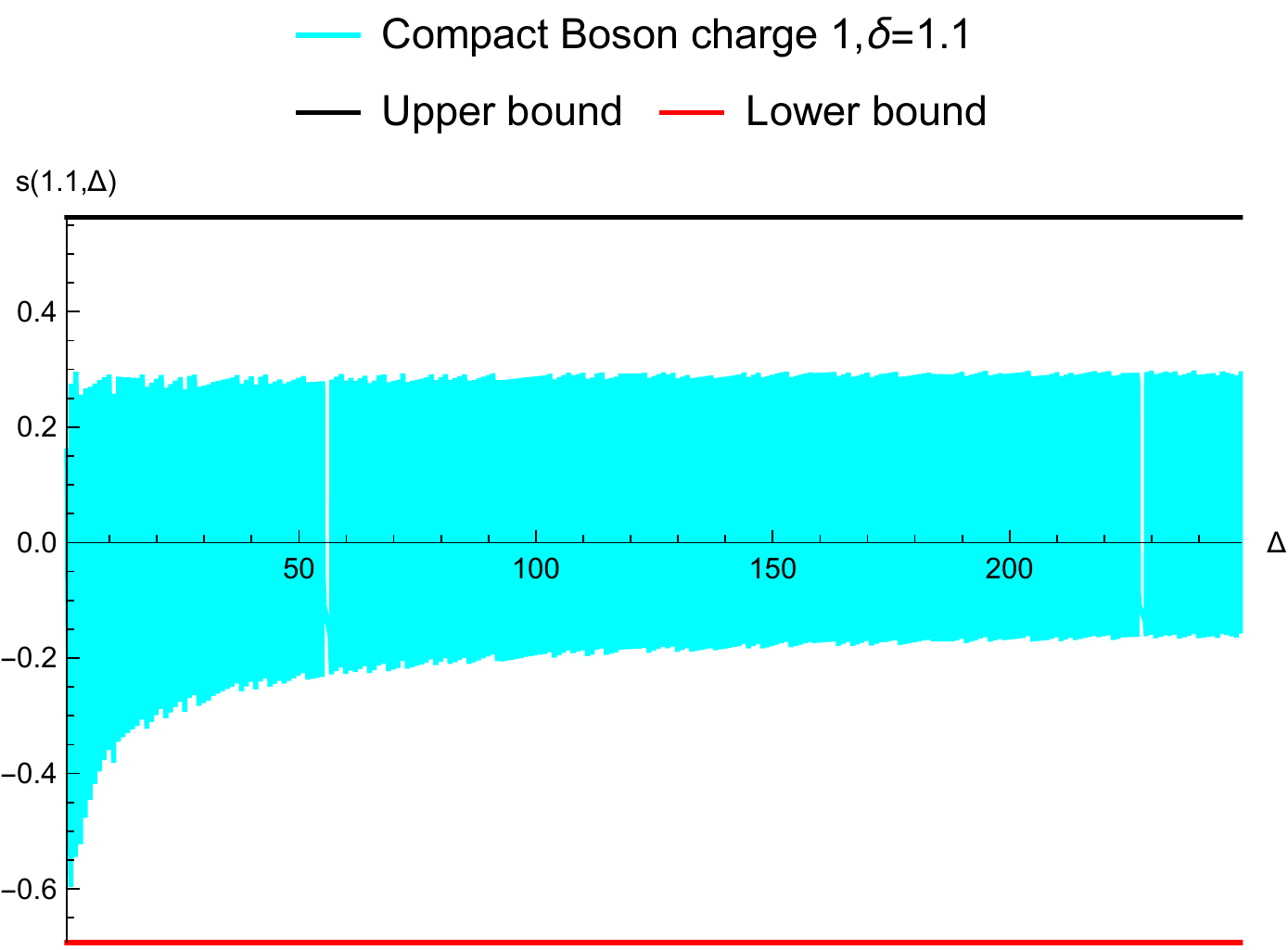}
\caption{Here, we estimate the number of operators with charge $Q=0,1$ corresponding to the $U(1)$ symmetry in compact boson at $R=\frac{1}{2}$. We plot the logarithm of the ratio of actual number of operators in the interval of size $2\delta=2.2$ and the leading prediction from Tauberian-Cardy analysis (in cyan). We see that they are well within the bounds (the black and the red line) as predicted in the main text.}
\label{cb}
\end{figure}

\section{Spin selection rule for anomalous symmetry}
\label{sec:spin}
The defect Hilbert space is defined by having a TDL along the time like direction. Now if we want to define the action of the symmetry in the defect Hilbert space, we need to introduce another TDL along the spatial direction. Since, the two TDLs cross each other, we need to resolve the crossing. And this is how the global symmetry can turn out to have 't Hooft anomaly, which is related to the ambiguity in locally resolving the crossing configuration of two TDL (see fig.~\ref{crossing-relation-anomaly}). Two different ways of resolution leads to defining two operators $\mathcal{L}_{\pm}$ acting on the states in the defect Hilbert space. Relationship between these two different ways of resolving ambiguity leads to the ``crossing relations'', which naturally generalize to the any TDLs (not only the one corresponding to the global symmetry). We will see that such 't Hooft anomaly of global symmetry will impose spin selection rules on the defect Hilbert space (see fig.~\ref{spin-selection-rule}).

\begin{figure}[H]
\centering
\includegraphics[scale = 0.2]{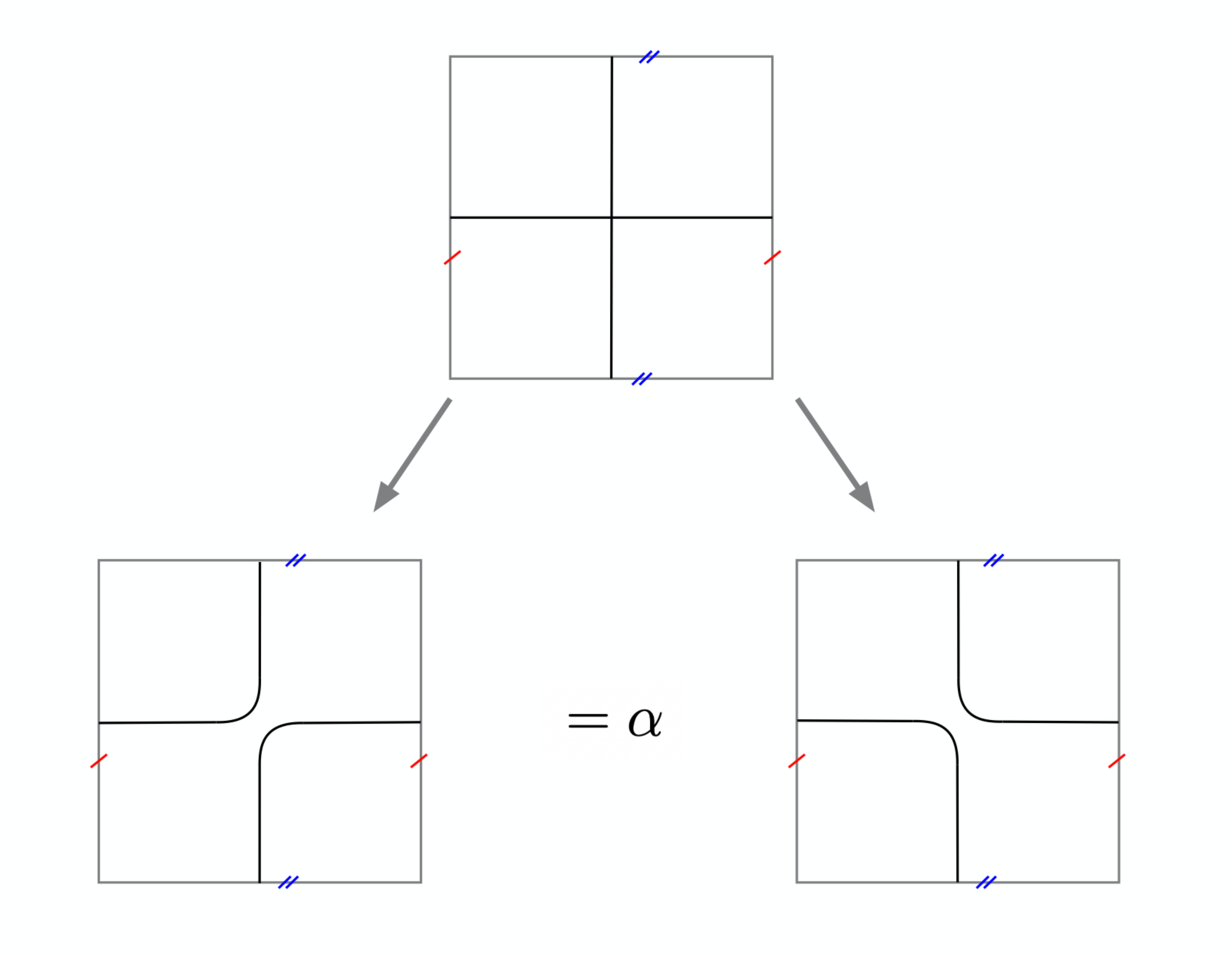}
\caption{Here, we consider the non-trivial line (black) generating $\mathbb{Z}_2$ symmetry on a torus. There are two ways to resolve the crossing configuration on the top, which are related by $\alpha = \pm 1$ where $1$ is for non-anomalous $\mathbb{Z}_2$ and $-1$ is for the anomalous $\mathbb{Z}_2$. We note the left configuration as $\hat{\mathcal{L}}_+$ and the right configuration as $\hat{\mathcal{L}}_-$.}
\label{crossing-relation-anomaly}
\end{figure}

We will focus on the group $\mathbb{Z}_2$ for rest of the appendix. Following \cite{Lin:2019kpn}, to derive a spin selection rule, we first determine the action of $\hat{\mathcal{L}}_{\pm}$ on the defect Hilbert space and then consider a specific configuration which relates the action of $\hat{\mathcal{L}}_{\pm}$ to the spin of the state. We consider the fig.~\ref{action-of-L-on-defect} to derive $\alpha^2=1$.
\begin{figure}[H]
\centering
\includegraphics[scale = 0.25]{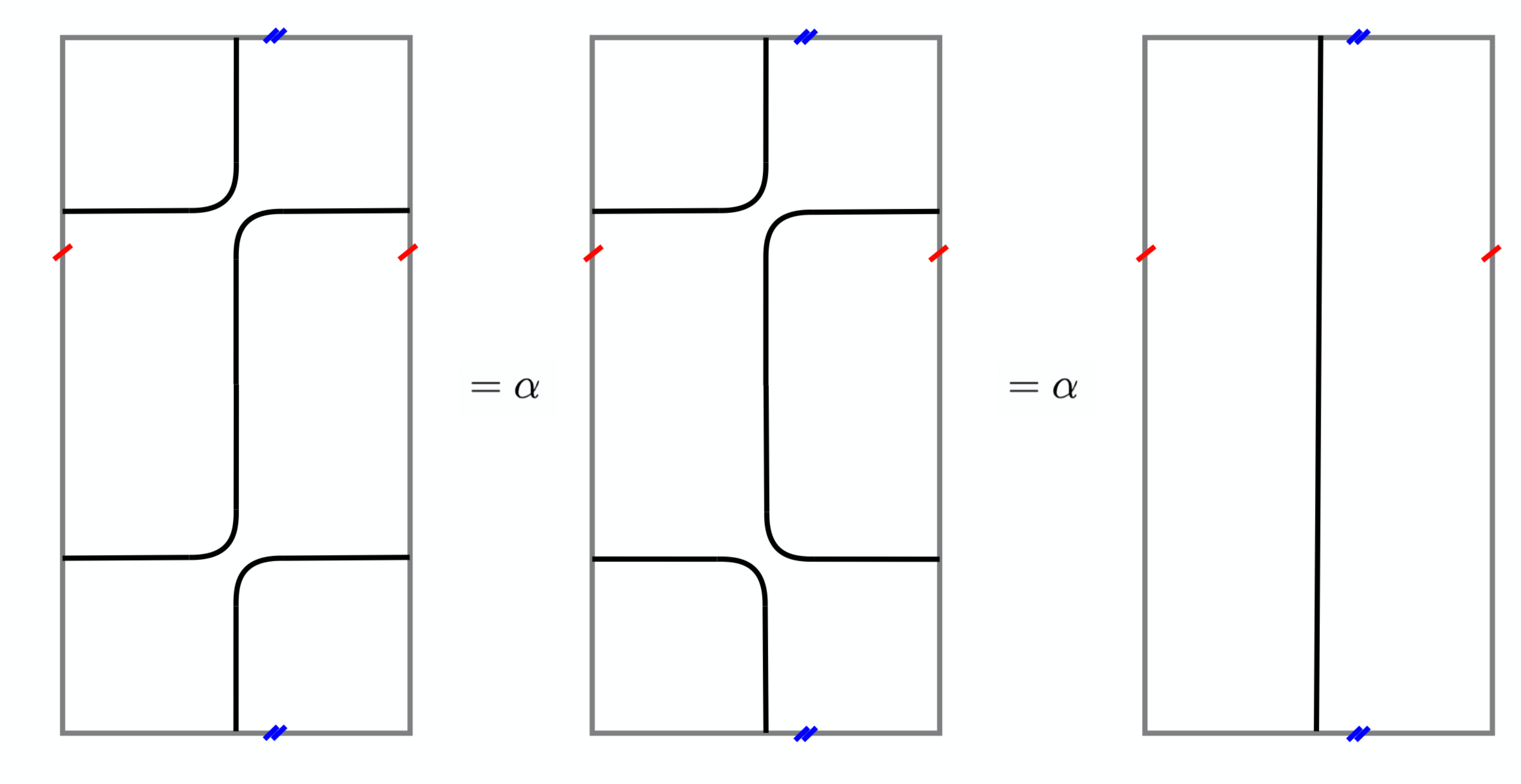}
\caption{Here we consider two $\hat{\mathcal{L}}_+$ on the left figure and show it is equal to $\alpha$ acting on states in the defect Hilbert space.}	
\label{action-of-L-on-defect}
\end{figure}
On the other hand, we have
\begin{equation}
\begin{aligned}
	(\hat{\mathcal{L}}_+)^2 |h,\overline{h}\rangle &= \alpha |h,\overline{h}\rangle
	\Rightarrow
	\hat{\mathcal{L}}_+ |h,\overline{h}\rangle &= \pm \sqrt{\alpha}|h,\overline{h}\rangle\,.
	\end{aligned}	
\end{equation}

For the next step, we consider mapping $\hat{\mathcal{L}}_+ |h,\oh\rangle$ from $\mathbb{R}_t\times S^1$ to $\mathbb{R}^2$ and unwind the $\mathcal{L}^+$ to deduce 
\begin{equation}
	\hat{\mathcal{L}}_{\pm}|h,\oh\rangle = e^{\pm 2\pi\imath s}|h,\oh\rangle.	
\end{equation}
\begin{figure}[H]
\centering
\includegraphics[scale = 0.25]{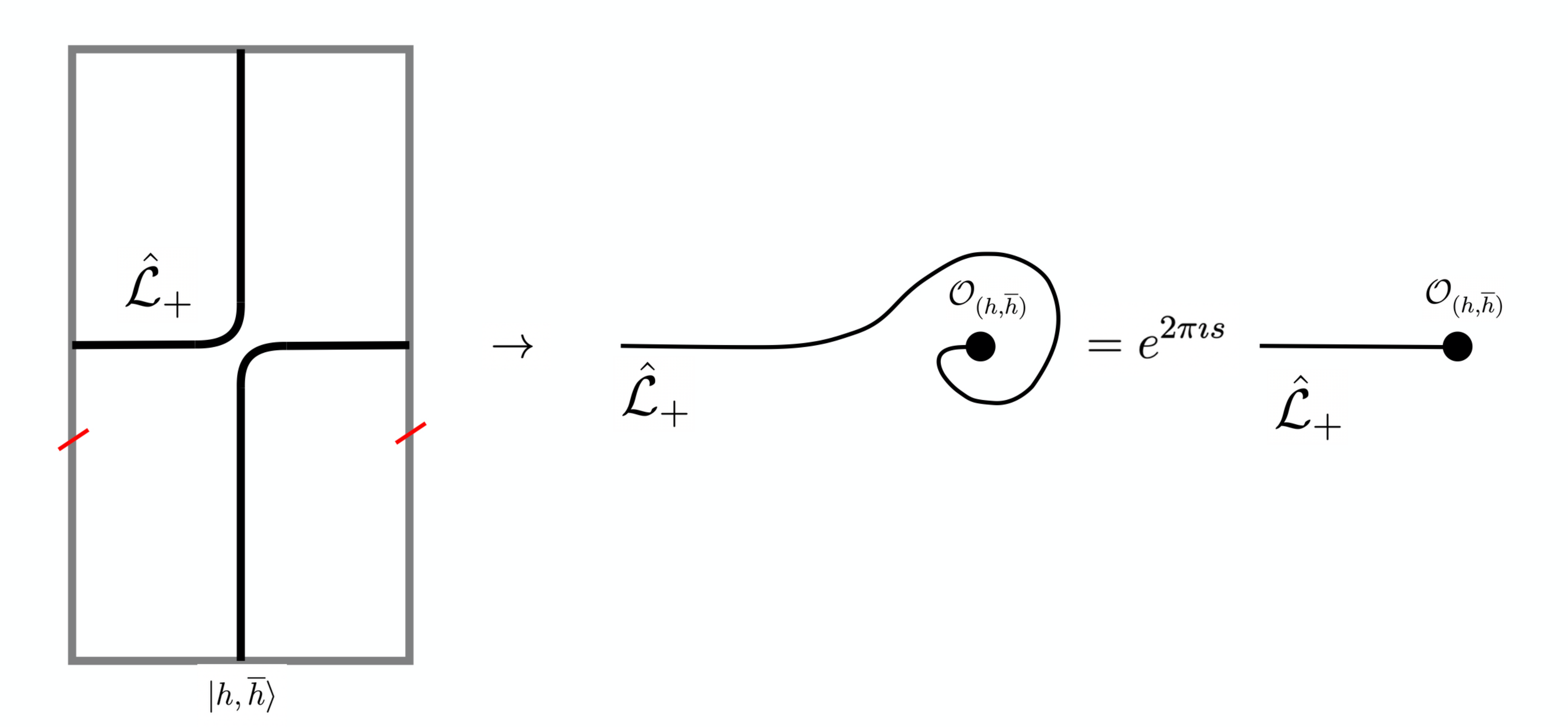}
\caption{Here, we consider the action of $\hat{\mathcal{L}}_+$ acting on the state $|h,\oh\rangle$ on $\mathbb{R}_t\times S^1$ and maps to $\mathbb{R}^2$ via the operator-state correspondence map. Then unwinding $\hat{\mathcal{L}}_+$ shows $\hat{\mathcal{L}}_+|h,\oh\rangle = e^{2\pi\imath s} |h,\oh\rangle$.}	
\label{spin-selection-rule}
\end{figure}

Combining the previous results, we find:
\begin{itemize}
	\item in the non-anomalous case where $\alpha = 1$, we have
	\begin{equation}
		s \in \begin{cases}
 		\mathbb{Z} \,\,\, &\text{if } \hat{\mathcal{L}}_+ \text{ acts as } + 1, \\
 		\frac{1}{2} + \mathbb{Z} \,\,\, &\text{if } \hat{\mathcal{L}}_+ \text{ acts as } -1;
 		\end{cases}
	\end{equation}
	\item in the anomalous case where $\alpha = -1$, we have
	\begin{equation}
		s \in \begin{cases}
 		+ \frac{1}{4} + \mathbb{Z} \,\,\, &\text{if } \hat{\mathcal{L}}_+ \text{ acts as } + i, \\
 		- \frac{1}{4} + \mathbb{Z} \,\,\, &\text{if } \hat{\mathcal{L}}_+ \text{ acts as } -i.
 		\end{cases}
	\end{equation}

\end{itemize}
Analogously, one can generalize the above result to $\mathbb{Z}_n$ \cite{Chang:2018iay}. Thus the spin selection rule automatically rules out the existence of $\Delta=0$ states in the defect Hilbert space if the symmetry is anomalous. For completeness, we remark here that if the symmetry is non-anomalous, we can rule out the existence of $\Delta=0$ state by requiring that that symmetry group acts faithfully on the Hilbert space.

\section{Review of Representation Theory for Finite Group}\label{srefreview}
We review some basic notions and results in the representation theory for finite group. For a more detailed exposition including proofs and jokes, see II.1 and II.2 of \cite{Zee:2016fuk}.\\

Given a finite group $G$ and a unitary (reducible or irreducible) representation $r$ of $G$ given by matrices $D^{(r)}(g)$, we define the character $\chi^{(r)}(g)$ to be
\begin{equation}
	\chi^{(r)}(g) \equiv \text{tr} D^{(r)}(g).
\end{equation}

The Great Orthogonality Theorem together with one of its corollary states that, given two irreducible representation $r,s$, 
\begin{equation}
	\sum_{g} D^{(r)\dagger}(g)^i{}_j D^{(s)}(g)^k{}_l = \frac{|G|}{d_r}\delta^{rs} \delta^i{}_l \delta^k{}_j	
\end{equation}
where $|G|$ is the order of the group, $d_r$ is the dimension of the irrep $r$, and $\delta^{rs} = 1$ if two irreps are the same and $\delta^{rs} = 0$ otherwise. For a proof of this result, see II.2 of \cite{Zee:2016fuk}.

From the above result, one can derive the so-called character orthogonality. By taking trace, we find
\begin{equation}
	\sum_g (\chi^{(r)}(g))^* \chi^{(s)}(g) = |G| \delta^{rs}.	
\end{equation}

We can use the character orthogonality to count how many times a given irrep $r$ appears in a reducible representation. First, notice that if a reducible representation $R$ can be decompose into a direct sum of irreps $r_i$, then 
\begin{equation}
	\chi^{(R)}(g) = \sum_{i=1}^{N} \chi^{(r_i)}(g).
\end{equation}

Applying the character orthogonality, we find
\begin{equation}
	\frac{1}{|G|}\sum_{g} (\chi^{(r)}(g))^* \chi^{(R)}(g)=\text{No of times irrep ``r'' appears}\,.
\end{equation}

In context of conformal field theory, the finite symmetry group $G$ commutes with the Virasoro algebra, thus the states with the same scaling dimension $\Delta$ form a reducible representation of $G$. Therefore,
\begin{equation}
\begin{aligned}
	\frac{1}{|G|}\sum_g \chi^\alpha(g)^* Z^{\mathcal{L}}(\beta,g) &= \frac{1}{|G|} \sum_g \chi^\alpha(g)^* \text{Tr}\bigg(\hat{g} q^{L_0 - c/24} \overline{q}^{\overline{L}_0 - c/24}\bigg) \\
	&= \frac{1}{|G|} \sum_g \sum_{\Delta} \chi^\alpha(g)^* (\text{Tr}_{H_\Delta} \hat{g}) e^{-\beta(\Delta-c/12)} \\
	&= \sum_{\Delta} N_{\alpha,\Delta} e^{-\beta(\Delta-c/12)}
\end{aligned}
\end{equation}
where $N_{\alpha,\Delta}$ is the number of irrep $\alpha$ with scaling dimension $\Delta$. We used this basic fact in the statements below \eqref{29}. }

\bibliography{refs}
\end{document}